\documentclass[12pt,epsf,amssymb]{article} \textheight =23 truecm
\def\lsim{\raise0.3ex\hbox{$<$\kern-0.75em\raise-1.1ex\hbox{$\sim$}}}
\def\gsim{\raise0.3ex\hbox{$>$\kern-0.75em\raise-1.1ex\hbox{$\sim$}}}
\def\noi{\noindent}  \def\bea{\begin{eqnarray}}
\def\eea{\end{eqnarray}} \def\beq{\begin{equation}}
\def\eeq{\end{equation}} 
\def\beeq{\begin{eqnarray}} \def\eeeq{\end{eqnarray}} \def\R{ {\rm R
\kern -.31cm I \kern .15cm}} \def\C{ {\rm C \kern -.15cm \vrule
width.5pt \kern .12cm}} \def\Z{ {\rm Z \kern -.27cm \angle \kern
.02cm}} \def\N{ {\rm N \kern -.26cm \vrule width.4pt \kern .10cm}}
\def\1{{\rm 1\mskip-4.5mu l} }

\usepackage{graphics} \usepackage{graphicx} \usepackage{epsfig}

\begin{document} 

\begin{center} 

{\Large \bf Lepton mixing in gauge models} \\

\vskip 1 truecm {\bf D. Falcone and L. Oliver}\\

\vskip 0.3 truecm

{\it Laboratoire de Physique Th\'eorique}\footnote{Unit\'e Mixte de
Recherche UMR 8627 - CNRS }\\    {\it Universit\'e de Paris XI,
B\^atiment 210, 91405 Orsay Cedex, France} 

\end{center}

\vskip 1 truecm

\begin{abstract}

We reexamine lepton mixing in gauge models by considering two theories within the type I seesaw mechanism, the Extended Standard Model, i.e. $SU(2)_L \times U(1)_{Y}$ with singlet right-handed heavy neutrinos, and the Left-Right Model $SU(2)_L \times SU(2)_R \times U(1)_{B-L}$. The former is often used as a simple heuristic approach to masses and mixing of light neutrinos and to leptogenesis, while we consider the latter as an introduction to other left-right symmetric gauge theories like $SO(10)$. We compare lepton mixing in both theories for general parameter space and discuss also some particular cases. In the electroweak broken phase, we study in parallel both models in the "current basis" (diagonal gauge interactions), and in the "mass basis" (diagonal mass matrices and mixing in the interaction), and perform the counting of $CP$ conserving and $CP$ violating parameters in both bases. We extend the analysis to the Pati-Salam model $SU(4)_C \times SU(2)_L \times SU(2)_R$ and to $SO(10)$. Although specifying the Higgs sector increases the predictive power, in the most general case one has the same parameter structure in the lepton sector for all the left-right symmetric gauge models. We make explicit the differences between the Extended Standard Model and the left-right models, in particular $CP$ violating and lepton-number violating new terms involving the $W_R$ gauge bosons. As expected, at low energy, the differences in the light neutrino spectrum and mixing appear only beyond leading order in the ratio of Dirac mass to right-handed Majorana mass. 

\end{abstract}

\vskip 0.5 truecm

\noi LPT-Orsay-14-41\par

\par \vskip 0.1 truecm

\noindent e-mails :  domenicofalcone3@virgilio.it, Luis.Oliver@th.u-psud.fr 

\newpage \pagestyle{plain}

\section{\bf Introduction}

In the last years, an impressive experimental progress has been achieved on the neutrino spectrum and mixing.
Using this information on the light neutrinos mass matrix $m_L$, one is tempted to use the inverse of the seesaw formula $M_R = -m_D^t m_L^{-1}m_D$, where $m_D$ is the Dirac neutrino mass matrix, as a window on high energy neutrino physics, i.e. on the heavy right-handed neutrino mass matrix $M_R$ \cite{F,AFS,AF,BFO,BFFNR}.\par 

To use the inverse seesaw formula one needs information on the crucial Dirac mass matrix $m_D$. It has been often suggested that theoretical information on this matrix can be guessed within the $SO(10)$ Grand Unification gauge theory \cite{SO10}.
In order to study the whole structure of $SO(10)$ as far as lepton mixing is concerned, we have realized that it is convenient to begin by considering simpler theories that also exhibit left-right (LR) symmetry (for a review, see ref. \cite{FALCONE}).\par 
The simplest gauge theory that has been builded to study lepton mixing is the one that we call Extended Standard Model (ESM), i.e. the Standard Model (SM) $SU(3) \times SU(2)_L \times U(1)_Y$ plus right-handed neutrinos $N_R$, one per generation, singlet under the SM gauge group. Although this scheme allows to introduce heavy right-handed neutrinos, it does not exhibit LR symmetry like $SO(10)$.\par 

One main aim of the present paper is to compare lepton mixing in the ESM, on the one hand, with lepton mixing in left-right models like $SO(10)$.
Lepton mixing in the ESM has been thoroughly studied in the literature \cite{SV,HMP,BMNR,BGFJR}, specially in ref. \cite{BMNR} on which the present paper heavily relies, together with the comprehensive review paper \cite{BGFJbis}. 

To compare the ESM with left-right gauge theories we have found convenient to consider next the Left-Right Model (LRM) $SU(2)_L \times SU(2)_R \times U_{B-L}(1)$ \cite{MP-1,MP-2}, that exhibits a number of interesting new features concerning lepton mixing \cite{LS-2,LS-1}. This gauge group has already an appreciable complexity that will be useful as an introduction for the study of larger LR gauge groups, like the Pati-Salam model $SU(4)_C \times SU(2)_L \times SU(2)_R$ \cite{PS}, and the grand unified $SO(10)$ gauge group \cite{SO10}.\par

We will first consider completely general Dirac or Majorana mass matrices consistent with Lorentz invariance, that coincide with mass matrices arising from the most general Higgs structure. We then look for the parameters that can be rotated away, although in a different way in the ESM and the LRM.
We will consider the {\it current basis}, in which the interaction Lagrangian ${\cal{L}}_w$ is diagonal, and the {\it mass basis}, in which the mass Lagrangian ${\cal{L}}_m$ is diagonal, and we check that, for a given model, the final number of independent parameters, angles and phases, is the same in both bases.\par

Some main results exposed below are already known. The purpose of this paper is in part didactic, and in part the understanding a number of particular points. We think it is worth to explain in detail the differences between the Extended Standard model and the Left-Right gauge models as far as lepton mixing is concerned, specially the comparison of the interaction Lagrangians of both schemes in the mass basis.\par

Here below we expose briefly the fermion and gauge boson content of the ESM and LRM. 
In Sections 2 and 3 we perform the counting of  the lepton sector parameters of the ESM and LRM in the current and in the mass bases. For the mass basis, special care is given to the approximation $m_D << M_R$, as compared with exact results, and in Section 4 we recall two different representations proposed in the literature  for the Dirac mass matrix $m_D$. In Section 5 we briefly examine leptogenesis in the ESM and in the LRM. In Section 6 we summarize the differences between both models for lepton mixing. Section 7 is devoted to the extension of our results to other left-right theories, Pati-Salam and $SO(10)$, and in Section 8 we conclude. In the Appendix we present some details of the calculations.\par 

\subsection{Gauge boson and fermion content of the gauge models}

We now expose the fermion and gauge boson content of the two gauge theories that we consider in detail, the Extended Standard model and the Left-Right model $SU(2)_L \times SU(2)_R \times U(1)_{B-L}$.

\vskip 0.5 truecm

\subsubsection{Extended Standard Model}

\vskip 0.3 truecm

The Extended Standard Model (ESM) is just the Standard Model (SM) $SU(3) \times SU(2)_L \times U(1)_Y$ with the addition of one Majorana fermion $N_R$ per generation, singlet under the gauge group.\par
The fermion content of the model is for quarks
\beq
\label{1.1e}  
\left(\begin{array}{c}
       u_L \\
         d_L \\
        \end{array}
        \right) \sim \left({\bf 3}, {\bf 2}, {1 \over 3} \right), \qquad \qquad
u_R \sim \left({\bf 3}, {\bf 1}, {4 \over 3} \right), \qquad \qquad d_R \sim \left({\bf 3}, {\bf 1}, -{2 \over 3} \right)
\eeq 

\noindent and for leptons
\beq
\label{1.2e}  
\left(\begin{array}{c}
       \nu_L \\
        e_L \\
        \end{array}
        \right) \sim \left({\bf 1}, {\bf 2}, -1 \right), \qquad \qquad  e_R \sim \left({\bf 1}, {\bf 1}, -2 \right), \qquad \qquad N_R \sim \left({\bf 1}, {\bf 1}, 0 \right)
\eeq

\noindent with 
\beq
\label{1.3e}  
Q = T_{3L} + {Y \over 2}
\eeq

The gauge bosons are the gluons $({\bf 8},{\bf 1},0)$, the $W_L$ bosons $({\bf 1},{\bf 3},0)$ and the $B$ boson $({\bf 1},{\bf 1},0)$.

The Higgs sector needed to achieve the Spontaneous Symmetry Breaking (SSB) and give masses to the fermions is the usual doublet $\phi \sim ({\bf 1},{\bf 2},-1)$.
The novelty in the ESM with respect to the SM is just the presence of the Majorana $N_R$ singlet. The right-handed fermion $N_R$ can have a large mass, of a different scale than the SM, that can be originated from a Higgs boson, singlet relatively to the Standard Model $\Phi \sim ({\bf 1},{\bf 1},0)$, or simply  be a bare mass term
\beq
\label{4.1bise} 
\left({\bf 1},{\bf 1},0 \right)_f \times \left({\bf 1},{\bf 1},0 \right)_f  = \left({\bf 1},{\bf 1},0 \right)
\eeq

\noindent that, together with the Dirac mass terms
\beq
\label{4.2bise} 
\left({\bf 1},{\bf 2},-1 \right)_f \times \left({\bf 1},{\bf 2},1 \right)_{\overline{f}} \times \left({\bf 1},{\bf 2},-1 \right)_H = \left({\bf 1},{\bf 1},0 \right) +\ ...
\eeq

\noindent gives the general neutrino mass matrix
\beq
\label{4.3e} 
\cal{M} = \left(\begin{array}{cc}
       0 & m_D \\
        m_D^t & M_R
        \end{array} \right)
\eeq

\noindent where $m_D$ and $M_R$ are respectively general complex and complex symmetric matrices.
 
\subsubsection{Left-Right Model}

\vskip 0.3 truecm

In the LRM model $SU(3) \times SU(2)_L \times SU(2)_R \times U(1)_{B-L}$, the classification of L and R fermions is for quarks
\beq
\label{1.4e}  
\left(\begin{array}{c}
       u_L \\
         d_L \\
        \end{array}
        \right) \sim \left({\bf 3}, {\bf 2}, {\bf 1}, {1 \over 3} \right), \qquad \qquad \left(\begin{array}{c}
       u_R \\
         d_R \\
        \end{array}
        \right) \sim \left({\bf 3}, {\bf 1}, {\bf 2}, {1 \over 3} \right)
\eeq 

\noindent and for leptons
\beq
\label{1.5e}  
\left(\begin{array}{c}
       \nu_L \\
        e_L \\
        \end{array}
        \right) \sim \left({\bf 1}, {\bf 2}, {\bf 1}, -1 \right), \qquad \qquad \left(\begin{array}{c}
       N_R \\
        e_R \\
        \end{array}
        \right) \sim \left({\bf 1}, {\bf 1}, {\bf 2}, -1 \right)  
\eeq

\noindent with 
\beq
\label{1.6e}  
Q = T_{3L} + T_{3R} + {B-L \over 2}
\eeq

The gauge bosons are the gluons $({\bf 8},{\bf 1},{\bf 1},0)$, the $W_L$ bosons $({\bf 1},{\bf 3},{\bf 1},0)$, the $W_R$ bosons $({\bf 1},{\bf 1},{\bf 3},0)$ and the $B-L$ singlet $({\bf 1},{\bf 1},{\bf 1},0)$. 

The Higgs fields needed to achieve SSB and the seesaw mechanism are the bidoublet $\phi \sim ({\bf 1},{\bf 2},{\bf 2},0)$ and the triplet $\Delta_R \sim ({\bf 1},{\bf 1},{\bf 3},2)$.\par 
The bidoublet, written as
\beq
\label{4.1e} 
\phi = \left(\begin{array}{cc}
       \phi_1^0 & \phi_1^+ \\
        \phi_2^- & \phi_2^0
        \end{array} \right)
\eeq

\noindent breaks the SM group and gives masses to quarks and leptons through the Yukawa terms
$$\left({\bf 3},{\bf 2},{\bf 1},{1 \over 3} \right)_f \times \left({\bf \overline{3}},{\bf 1},{\bf 2},-{1 \over 3} \right)_{\overline{f}} \times \left({\bf 1},{\bf 2},{\bf 2}, 0 \right)_{H,\overline{H}} = \left({\bf 1},{\bf 1},{\bf 1},0 \right) +\ ...$$
\beq
\label{4.2e} 
\left({\bf 1},{\bf 2},{\bf 1},-1 \right)_f \times \left({\bf 1},{\bf 1},{\bf 2},1 \right)_{\overline{f}} \times \left({\bf 1},{\bf 2},{\bf 2}, 0 \right)_{H,\overline{H}} = \left({\bf 1},{\bf 1},{\bf 1},0 \right) +\ ...
\eeq 

\noindent with $H = \phi$ and $\overline{H} = \sigma_2 H^* \sigma_2$.\par
From the vacuum expectation values
\beq
\label{4.3bise} 
<\phi_1^0>\ = k_1, \qquad \qquad \qquad <\phi_2^0>\ = k_2
\eeq

\noindent that can be complex, the Yukawa couplings give the Dirac masses, as in the SM, but with a different pattern. Quark mass matrices $m_u$, $m_d$ and the Dirac neutrino mass matrix $m_D$ read
$$m_u = p k_1 + q k_2^*, \qquad \qquad \qquad m_d = p k_2 + q k_1^*$$
\beq
\label{4.4e} 
m_D = r k_1 + s k_2^*, \qquad \qquad \qquad m_e = r k_2 + s k_1^*
\eeq

\noindent where $p$, $q$, $r$ and $s$ are complex Yukawa coupling matrices.\par
The triplet $H = \Delta_R$ breaks the LR model to the SM and, at the same time, gives a Majorana mass to the right-handed neutrino $N_R$ through the Yukawa term
\beq
\label{4.5e} 
\left({\bf 1},{\bf 1},{\bf 2},-1 \right)_f \times \left({\bf 1},{\bf 1},{\bf 2},-1 \right)_f \times \left({\bf 1},{\bf 1},{\bf 3}, 2 \right)_H = \left({\bf 1},{\bf 1},{\bf 1},0 \right) +\ ...
\eeq
\beq
\label{4.6e} 
<\Delta^0_R>\ = v_R, \qquad \qquad M_R = t v_R, \qquad \qquad M_R^t = M_R
\eeq

\noindent where $t$ is a complex symmetric Yukawa coupling matrix.

 The full neutrino mass matrix has the form
\beq
\label{4.6bise} 
\cal{M} = \left(\begin{array}{cc}
       0 & r k_1 + s k_2^* \\
        r^t k_1 + s^t k_2^* & t v_R
        \end{array} \right)
\eeq

\noindent i.e. it has the general form (\ref{4.3e}).\par
We consider this minimal Higgs content that is necessary in the LRM, and we do not introduce a possible left-handed triplet $\Delta_L = \left({\bf 1},{\bf 3},{\bf 1}, 2 \right)_H$ that could in principle contribute to the light neutrino masses.

\section{\bf Current basis}

In what follows, we consider the gauge models in the {\it electroweak broken phase}. We only make explicit the charged current terms in the interaction Lagragians of both gauge models.

\subsection{\bf Extended Standard Model}

The mass and interaction Lagrangians write, in an obvious compact notation
$${\cal{L}}_m = \overline{\nu}_L m_D N_R + {1 \over 2}\ \overline{(N_R)^c}M_RN_R + \overline{e}_L m_e e_R + h.c.$$
\beq
\label{2.1e}
{\cal{L}}_w = \overline{\nu}_L\gamma_\mu e_L W_L^{\mu} + h.c. \qquad \qquad \qquad \qquad \qquad \qquad
\eeq

\noindent The matrices $m_D$ and $m_e$ are general complex, each has 9 complex parameters, while
$M_R$ is general complex symmetric with 6 complex parameters.\par
The lepton number assignment $L(N_R) = - L((N_R)^c) = 1$ implies that the Majorana mass term is $\mid \Delta L\mid\ = 2$ while, like for the other fermions, while the Dirac mass term is $\mid \Delta L\mid\ = 0$.

From now on we adopt the following simplifying notation for the real parameters of an arbitrary square complex matrix $M$, that has $n(m)$ parameters, where $n$ is the total number of real parameters, among which there are $m\  (m \leq n)$ are phases : 
\beq
\label{2.1bise}
\fbox{ ${M\ \rm {has}\ n(m)\
real\ parameters}\ \ \ \leftrightarrow \ \ \ \rm{n\ real\ parameters,\ m \leq n\ phases}$ }
\eeq

In this example, $m_D$ and $m_e$ have 18(9) real parameters and $M_R$ has 12(6) real parameters. Therefore, {\it a priori} one has in this model 30(15) real parameters.\par

Let us see now that we can reduce the number of independent parameters without modifying the interaction Lagrangian ${\cal{L}}_w$. Diagonalizing $m_e$ and $M_R$ by
\beq
\label{2.2e}
m_e = V_{eL}^\dagger m_e^{diag} V_{eR}, \qquad \qquad \qquad M_R = U_R^t M_R^{diag} U_R
\eeq

\noindent and redefining the fields
\beq
\label{2.3e}
U_RN_R \to N_R, \qquad \qquad V_{eR}e_R \to e_R, \qquad \qquad 
	\left(\begin{array}{c}
       V_{eL}\nu_L \\
        V_{eL}e_L \\
        \end{array}
        \right) \to \left(\begin{array}{c}
       \nu_L \\
        e_L \\
        \end{array}
        \right)
\eeq

\noindent one gets
$${\cal{L}}_m = \overline{\nu}_L V_{eL}m_DU_R^\dagger N_R + {1 \over 2}\ \overline{(N_R)^c}M_R^{diag}N_R + \overline{e}_L m_e^{diag} e_R + h.c. \qquad $$
\beq
\label{2.4e}
{\cal{L}}_w = \overline{\nu}_L \gamma_\mu e_LW_L^\mu + h.c. \ \qquad \qquad \qquad \qquad \qquad \qquad \qquad \qquad \qquad 
\eeq

\noindent The simultaneous transformation of $\nu_L$ and $e_L$ in (\ref{2.3e},\ref{2.4e}) ensures the invariance of ${\cal{L}}_w$, but then $V_{eL}$ appears in the Dirac mass term. Since $m_D$ is a general complex symmetric matrix, so is $V_{eL}m_DU_R^\dagger$.
Changing the notation
\beq
\label{2.5e}
V_{eL}m_DU_R^\dagger \to m_D \qquad \qquad \qquad
\eeq

\noindent one obtains
$${\cal{L}}_m = \overline{\nu}_L m_D N_R + {1 \over 2}\ \overline{(N_R)^c}M_R^{diag}N_R + \overline{e}_L m_e^{diag} e_R + h.c. $$
\beq
\label{2.6e}
{\cal{L}}_w = \overline{\nu}_L \gamma_\mu e_LW_L^\mu + h.c. \ \ \qquad \qquad \qquad \qquad \qquad \qquad \ \ \  
\eeq

We can redefine the doublet $\left(\begin{array}{c}
       \nu_L \\
        e_L \\
        \end{array}
        \right)$ and the singlet $e_R$ by the same diagonal phase matrix $P_e$ :
\beq
\label{2.7e}
\left(\begin{array}{c}
       \nu_L \\
        e_L \\
        \end{array}
        \right) \to \left(\begin{array}{c}
       P_e\nu_L \\
        P_ee_L \\
        \end{array}
        \right), \qquad \qquad \qquad e_R \to P_e e_R  
\eeq

\noindent and one gets
$${\cal{L}}_m = \overline{\nu}_L P^*_e m_D N_R + {1 \over 2}\ \overline{(N_R)^c}M_R^{diag}N_R + \overline{e}_L m_e^{diag} e_R + h.c.$$
\beq
\label{2.8e}
{\cal{L}}_w = \overline{\nu}_L \gamma_\mu e_LW_L^\mu + h.c. \ \ \qquad \qquad \qquad \qquad \qquad \qquad \qquad 
\eeq

\noindent Finally we can choose the phase matrix $P_e$ to cancel three phases of $m_D$ in $P^*_e m_D$ :
$${\cal{L}}_m = \overline{\nu}_L m_D N_R + {1 \over 2}\ \overline{(N_R)^c}M_R^{diag}N_R + \overline{e}_L m_e^{diag} e_R + h.c.$$
\beq
\label{2.9e}
{\cal{L}}_w = \overline{\nu}_L \gamma_\mu e_LW_L^\mu + h.c. \ \ \ \ \ \qquad \qquad \qquad \qquad \qquad \qquad
\eeq

\noindent where now the Dirac mass matrix $m_D$ is {\it not} a general complex matrix, but has 9 real parameters + 6 phases, i.e. 15(6) real parameters.\par

To summarize parameter counting, one is left in the current basis with 15(6) (from $m_D$) + 3(0) (from $m_e^{diag}$) + 3(0) (from $M_R^{diag}$) = 21(6) real parameters, i.e. among them 6 phases. This counting agrees with the one performed in ref. \cite{BGJ}. 

\vskip 1.0 truecm

\subsection{\bf Left-Right Model}

In the LRM, the Lagrangian in the lepton sector writes 
$${\cal{L}}_m = \overline{\nu}_L m_D N_R + {1 \over 2}\ \overline{(N_R)^c}M_RN_R + \overline{e}_L m_e e_R + h.c.$$
\beq
\label{2.10e}
{\cal{L}}_w = \overline{\nu}_L \gamma_\mu e_L W_L^\mu + \overline{N}_R \gamma_\mu e_R W_R^\mu + h.c. \ \ \ \ \qquad \qquad
\eeq

\noindent Notice that, to simplify the notation, possible $W_L-W_R$ mixing is for the moment neglected in the interaction term, that will be considered later. The matrices $m_D$ and $m_e$ are a priori general complex with 18(9) parameters each, and
$M_R$ is a general complex symmetric matrix with 12(6) parameters.\par
An important remark is in order here. Parameter counting of the Left-Right Model in the "Current basis" means that we are assuming the whole interaction Lagrangian ${\cal{L}}_w$ in (\ref{2.10e}) to be diagonal, both in the left and the right sectors. For low energy neutrino physics, it can seem academic to assume that the right-handed piece $\overline{N}_R \gamma_\mu e_R W_R^\mu + h.c.$ is kept diagonal, because it is an interaction term involving high scale degrees of freedom. However, this natural assumption in any LR gauge theory is not only a formal point since, to keep this piece diagonal amounts to assume that one assigns a lepton number to the $N_R$ neutrinos, just in the same way as it is done for the $\nu_L$ neutrinos in (\ref{2.10e}), and in consistency with the assignment $L(N_R) = - L((N_R)^c) = 1$ in the ESM. As we will see below, the diagonalization of the light neutrino mass matrix and of the right neutrino mass matrix will result in mixing matrices of the PMNS type for both the light and the heavy neutrinos. 

Diagonalizing $m_e$ by (\ref{2.2e}) and redefining the fields
\beq
\label{2.11e} 
	\left(\begin{array}{c}
       V_{eL}\nu_L \\
        V_{eL}e_L \\
        \end{array}
        \right) \to \left(\begin{array}{c}
       \nu_L \\
        e_L \\
        \end{array}
        \right),\qquad \qquad \left(\begin{array}{c}
       V_{eR}N_R \\
        V_{eR}e_R \\
        \end{array}
        \right) \to \left(\begin{array}{c}
       N_R \\
        e_R \\
        \end{array}
        \right) 
\eeq

\noindent one gets
$$\qquad \qquad {\cal{L}}_m = \overline{\nu}_L V_{eL}m_D V_{eR}^\dagger N_R + {1 \over 2}\ \overline{(N_R)^c}V_{eR}^*M_RV_{eR}^\dagger N_R + \overline{e}_L m_e^{diag} e_R + h.c.$$
\beq
\label{2.12e}
\qquad \qquad \ \ {\cal{L}}_w = \overline{\nu}_L \gamma_\mu  e_L W_L^\mu + \overline{N}_R \gamma_\mu e_R W_R^\mu + h.c. \ \ \ \qquad \qquad \qquad \qquad \qquad
\eeq

\noindent Since $m_D$ is general complex, so is $V_{eL} m_D V_{eR}^\dagger$, and $M_R$ being complex symmetric, so is $V_{eR}^*M_RV_{eR}^\dagger$. \par
Changing the notation
\beq
\label{2.13e}
V_{eL} m_D V_{eR}^\dagger \to m_D, \qquad \qquad \qquad V_{eR}^*M_RV_{eR}^\dagger \to M_R
\eeq

\noindent one obtains
$${\cal{L}}_m = \overline{\nu}_L m_D N_R + {1 \over 2}\ \overline{(N_R)^c}M_RN_R + \overline{e}_L m_e^{diag} e_R + h.c.$$
\beq
\label{2.14e}
{\cal{L}}_w = \overline{\nu}_L \gamma_\mu e_L W_L^\mu + \overline{N}_R \gamma_\mu e_R W_R^\mu + h.c. \qquad \qquad \ \ \ \ \ \  
\eeq

\noindent We can redefine the doublets by the same diagonal phase matrix $P_e$ :
\beq
\label{2.15e}
\left(\begin{array}{c}
       \nu_L \\
        e_L \\
        \end{array}
        \right) \to \left(\begin{array}{c}
       P_e\nu_L \\
        P_ee_L \\
        \end{array}
        \right), \qquad \ \ \ \left(\begin{array}{c}
       N_R \\
        e_R \\
        \end{array}
        \right) \to \left(\begin{array}{c}
       P_eN_R \\
        P_ee_R \\
        \end{array}
        \right)  
\eeq

\noindent and one gets
$${\cal{L}}_m = \overline{\nu}_L P^*_e m_D P_e N_R + {1 \over 2}\ \overline{(N_R)^c}P_eM_RP_eN_R + \overline{e}_L m_e^{diag} e_R + h.c.$$
\beq
\label{2.16e}
{\cal{L}}_w = \overline{\nu}_L \gamma_\mu e_L W_L^\mu + \overline{N}_R \gamma_\mu e_R W_R^\mu + h.c. \ \qquad \qquad \qquad \qquad \ \ \ \ \ \
\eeq

\noindent We can chose the phase matrix $P_e$ to cancel three phases of $m_D$ or three phases of $M_R$, but not both at the same time. We choose to absorb 3 phases in $M_R$. Changing the notation $P^*_e m_D P_e \to m_D$, one gets finally 
$${\cal{L}}_m = \overline{\nu}_L m_D N_R + {1 \over 2}\ \overline{(N_R)^c}M_RN_R + \overline{e}_L m_e^{diag} e_R + h.c.$$
\beq
\label{2.17e}
\ {\cal{L}}_w = \overline{\nu}_L \gamma_\mu e_L W_L^\mu + \overline{N}_R \gamma_\mu e_R W_R^\mu + h.c. \ \ \qquad \qquad \qquad   
\eeq

\noindent where $m_D$ is an arbitrary complex matrix with 18(9) parameters and $M_R$ is complex symmetric with 9(3) parameters.\par
To summarize, one gets finally in the LRM : 18(9) parameters from $m_D$ + 9(3) parameters from $M_R$ + 3 eigenvalues in $m_e^{diag}$ = 30(12) parameters.\par
Much more constrained models have been considered in the literature. For example, the {\it Minimal} LRM within supersymmetry with a Higgs content that implies $m_e = m_D, m_u = m_d$ (up-down unification) \cite{BDM}, that has a reduced number of parameters. 

\section{\bf Mass basis}

\subsection{\bf Extended Standard Model}

\vskip 0.3 truecm

For the diagonalization of the whole $6 \times 6$ neutrino mass matrix, we proceed step by step, and we begin with (\ref{2.9e}), where $m_e^{diag}$ and $M_R^{diag}$ are diagonal and the Dirac mass matrix $m_D$ has 15(6) parameters.
So we can rewrite
$${\cal{L}}_m = {1 \over 2} \left(\overline{\nu_L}, \ \overline{(N_R)^c} \right) {\cal{M}} \left(\begin{array}{c}

       \nu_L^c \\
       N_R\\
        \end{array}
        \right) + \overline{e}_L m_e^{diag} e_R + h.c. \qquad \ $$
\beq
\label{3.1.2e}
{\cal{L}}_w = \overline{\nu}_L \gamma_\mu e_L W_L^\mu + h.c. \ \ \ \ \ \ \qquad \qquad \qquad \qquad \qquad \qquad \ \ \  
\eeq

\noindent where $\cal{M}$ has the form
\beq
\label{3.1.3e}
\cal{M} = \left(
        \begin{array}{ccc}
       0 & m_D \\
        m_D^t & M_R^{diag} \\
  	\end{array}
        \right) 
\eeq

\noindent This matrix has 18(6) parameters : 15(6) from $m_D$ and 3(0) from $M_R^{diag}$.\par
Let us now diagonalize $\cal{M}$ with the unitary matrix $V$ \cite{HMP,BMNR,BGFJR}
\beq
\label{3.1.4e}
{\cal{M}} = V {\cal{M}}^{diag} V^t 
\eeq

\noindent where
\beq
\label{3.1.5e}
{\cal{M}}^{diag} = \left(\begin{array}{ccc}
       m_L^{diag} & 0 \\
        0 & M_R^{diag} \\
  	\end{array}
        \right)
\eeq
\beq
\label{3.1.6e}
V = \left(\begin{array}{ccc}
       K & R \\
        S & T \\
  	\end{array}
        \right) 
\eeq

\noindent Notice that since ${\cal{M}}^{diag}$ has 6 eigenvalues, and ${\cal{M}}$ has 18(6) parameters, the $6 \times 6$ unitary matrix $V$ will have 18(6) - 6(0 = 12(6) parameters.
Rewriting (\ref{3.1.2e}) under the form
$${\cal{L}}_m = {1 \over 2} \left(\overline{\nu_L}, \ \overline{(N_R)^c} \right) V {\cal{M}}^{diag} V^t \left(\begin{array}{c}
       \nu_L^c \\
       N_R\\
        \end{array}
        \right) + \overline{e}_L m_e^{diag} e_R + h.c. \qquad \ $$
\beq
\label{3.1.7e}
{\cal{L}}_w = \left(\overline{\nu_L}, \ \overline{(N_R)^c} \right) \gamma_\mu \left(\begin{array}{cc}
       1 & 0 \\
        0 & 0 \\
  	\end{array}
        \right) \left(\begin{array}{c}
       e_L \\
       e_L\\
        \end{array}\right)W_L^\mu + h.c. \ \ \qquad \qquad \qquad 
\eeq

\noindent and redefining
\beq
\label{3.1.8e}
V^t \left(\begin{array}{c}
       \nu_L^c \\
       N_R\\
        \end{array}
        \right) \to \left(\begin{array}{c}
       \nu_L^c \\
       N_R\\
        \end{array}
        \right), \qquad \ \ \ \left(\overline{\nu}, \ \overline{(N_R)^c} \right) V \to \left(\overline{\nu}, \ \overline{(N_R)^c} \right) 
\eeq

\noindent one gets
$${\cal{L}}_m = {1 \over 2} \left(\overline{\nu_L}, \ \overline{(N_R)^c} \right) {\cal{M}}^{diag} \left(\begin{array}{c}
       \nu_L^c \\
       N_R\\
        \end{array}
        \right) + \overline{e}_L m_e^{diag} e_R + h.c. \qquad \ $$
\beq
\label{3.1.9e}
{\cal{L}}_w = \left(\overline{\nu_L}, \ \overline{(N_R)^c} \right) \gamma_\mu V^\dagger \left(\begin{array}{cc}
       1 & 0 \\
        0 & 0 \\
  	\end{array}
        \right) \left(\begin{array}{c}
       e_L \\
       e_L\\
        \end{array}\right)W_L^\mu + h.c. \ \ \ \ \ \qquad 
\eeq

\noindent or
$${\cal{L}}_m = {1 \over 2}\ \overline{\nu_L}m_L^{diag} (\nu_L)^c + {1 \over 2}\ \overline{(N_R)^c}M_R^{diag} N_R  + \overline{e}_L m_e^{diag} e_R + h.c. \qquad $$
\beq
\label{3.1.11e}
\fbox{ ${\cal{L}}_w = \left(\overline{\nu_L} K^\dagger e_L + \overline{(N_R)^c} R^\dagger e_L\right) \gamma_\mu W_L^\mu + h.c.$ } \ \qquad \qquad \qquad \qquad \ \ \
\eeq

\noindent The first term in ${\cal{L}}_w$ describes the Standard Model $\Delta L = 0$ decay
\beq
\label{3.1.11bise}
W_L \to e_L \overline{\nu}_L
\eeq

\noindent while the second term corresponds to the well-known $\Delta L = 2$ process
\beq
\label{3.1.11tere}
(N_R)^c \to e_L W_L
\eeq

\noindent ($L(N_R) = - L((N_R)^c) = L(e_L)$). The notation $(N_R)^c$ for the heavy neutrino makes explicit also the chirality conservation of the $V-A$ interaction.\par 

Notice that only the $3 \times 3$ complex matrices $K$ and $R$ from the $6 \times 6$ unitary matrix (\ref{3.1.6e}) are involved in the formula (\ref{3.1.11e}). Let us now count the parameters of these matrices.
From the zero in the matrix $\cal{M}$ (\ref{3.1.3e}) and the definitions (\ref{3.1.4e}-\ref{3.1.6e}) one finds (see eqn. (\ref{3.1.27e}) of the Appendix for $m_L = 0$)
\beq
\label{3.1.12e}
Km_L^{diag}K^t + RM_R^{diag}R^t = 0  
\eeq

\noindent Using the unitarity of the matrix $V$ (\ref{3.1.6e}) one has
\beq
\label{3.1.13e}
KK^\dagger + RR^\dagger = 1
\eeq

Eqns. (\ref{3.1.12e}) and (\ref{3.1.13e}) are identities between $3 \times 3$ matrices involving only the mixing matrices $K$ and $R$ and not the whole matrix (\ref{3.1.6e}). Due to these relations, {\it the matrices $K$ and $R$ are correlated}.\par

The conditions (\ref{3.1.12e}) and (\ref{3.1.13e}) reduce the number of independent parameters. Equation (\ref{3.1.12e}) is self-transposed, and gives 12(6) constraints, while (\ref{3.1.13e}) is hermitian, giving 9(3) constraints. This reduces the number of parameters of the two complex matrices $K$ and $R$ from 36(18) down to 15(9).\par
Finally, redefining the charged lepton fields by a diagonal $3 \times 3$ phase matrix $Q_e$ 
\beq
\label{3.1.14e}
e_L \to Q_e^\dagger e_L, \qquad  \qquad \qquad e_R \to Q_e^\dagger e_R
\eeq

\noindent one gets, from (\ref{3.1.11e}),
$${\cal{L}}_m = {1 \over 2}\ \overline{\nu_L}m_L^{diag} (\nu_L)^c + {1 \over 2}\ \overline{(N_R)^c}M_R^{diag} N_R  + \overline{e}_L m_e^{diag} e_R + h.c. \qquad \ \ $$
\beq
\label{3.1.15e}
{\cal{L}}_w = \left(\overline{\nu_L} (Q_eK)^\dagger + \overline{(N_R)^c}  (Q_eR)^\dagger \right) \gamma_\mu e_L W_L^\mu + h.c  \qquad \qquad \qquad \qquad 
\eeq

\noi On the other hand, multiplying (\ref{3.1.12e}) on the left by $Q_e$ and on the right by $Q_e^t$, and (\ref{3.1.13e}) on the left by $Q_e$ and on the right by $Q_e^\dagger$, these equations become
\beq
\label{3.16e}
(Q_eK)m_L^{diag}(Q_eK)^t + (Q_eR)M_R^{diag}(Q_eR)^t = 0  
\eeq
\beq
\label{3.1.17e}
(Q_eK)(Q_eK)^\dagger + (Q_eR)(Q_eR)^\dagger = 1
\eeq

\noindent and we can absorb 3 phases of one of the matrices $K$ or $R$, but not of both matrices at the same time.\par

In summary, the matrices $K$ and $R$ have {\it together} 12(6) parameters, and adding the 9(0) parameters from $m_e^{diag}$, $m_L^{diag}$ and $M_R^{diag}$ one obtains a total of 21(6) parameters, the same number as in the current basis.
In the ESM the matrices $K$ and $R$ are {\it decoupled} from $S$ and $T$ of (\ref{3.1.6e}), and obey relations (\ref{3.1.12e},\ref{3.1.13e}).\par  
We can now go somewhat further by considering first the whole matrix (\ref{3.1.6e}), and assuming $m_D << M_R$.\par

\subsubsection {The matrices $K$, $R$, $S$, $T$ in the Extended Standard Model}

\vskip 3truemm

Starting from the Lagrangian in the current basis (\ref{2.9e}), $m_D$ has now 15(6) parameters. Particularizing formulas (\ref{3.1.27e}-\ref{3.1.29e}) of the Appendix to the present case, we have :
\beq
\label{3.1.30e}
Km_L^{diag}K^t + RM_R^{diag}R^t = 0 \ \ \ \  
\eeq 
\beq
\label{3.1.31e}
Sm_L^{diag}S^t + TM_R^{diag}T^t = M_R^{diag}
\eeq
\beq
\label{3.1.32e}
Km_L^{diag}S^t + RM_R^{diag}T^t = m_D \ \ 
\eeq

Considering for the moment the unitarity of the matrix (\ref{3.1.6e}), the number of independent parameters in the l.h.s. will be 36(21) from $(K, R, S, T)$ + 3(0) from $m_L^{diag}$ + 3(0) from $M_R^{diag}$ = 42(21).\par

The complex symmetric matrix equation (\ref{3.1.30e}) gives 12(6) constraints. On the other hand, $M_R^{diag}$ appears already in the r.h.s. of (\ref{3.1.31e}), and this equation implies 12(6) - 3(0) = 9(6) constraints. Since $m_D$ has now 15(6) free parameters, eq. (\ref{3.1.32e}) gives 3(3) constraints, giving a total of 12(6) + 9(6) + 3(3) = 24(15) constraints. Therefore the number of independent parameters is 42(21) - 24(15) = 18(6) parameters. Adding the 3(0) eigenvalues of $m_e^{diag}$ one gets 18(6) + 3(0) = 21(6) parameters, the same result as in the current basis. 

Moreover, substracting from this total of 21(6) parameters the 9(0) mass eigenvalues $m_e^{diag}$, $m_L^{diag}$ and $M_R^{diag}$, the set of matrices $(K, R, S, T)$ has 12(6) parameters, the same number that we have found for $K$ and $R$, so that $S$ and $T$ are not independent.\par

\vskip 3truemm

\underline {Exact relations between the matrices $K$, $R$, $S$, $T$} 

\vskip 3truemm

On the other hand, from (\ref{3.1.3e}-\ref{3.1.6e}) one has

\beq
\label{3.1.33e}
\left(\begin{array}{ccc}
       0 & m_D \\
        m_D^t & M_R^{diag} \\
  	\end{array}
        \right) \left(\begin{array}{ccc}
       K^* & R^* \\
        S^* & T^* \\
  	\end{array}
        \right) = \left(\begin{array}{ccc}
       K & R \\
        S & T \\
  	\end{array}
        \right) \left(\begin{array}{ccc}
       m_L^{diag} & 0 \\
        0 & M_R^{diag} \\
  	\end{array}
        \right)
\eeq

\noindent hence
\beq
\label{3.1.34e}
\left(\begin{array}{ccc}
       m_D S^* & m_D T^* \\
       m_D^t K^*+M_R^{diag}S^* & m_D^t R^*+M_R^{diag}T^* \\
  	\end{array}
        \right) = \left(\begin{array}{ccc}
       Km_L^{diag} & RM_R^{diag} \\
        Sm_L^{diag} & TM_R^{diag} \\
  	\end{array}
        \right)
\eeq

\noindent and therefore one obtains the following exact expressions of the matrices $R$, $S$ in terms of $K$, $T$, $m_D$ and the mass eigenvalues :
\beq
\label{3.1.35e}
R = m_D T^*(M_R^{diag})^{-1}
\eeq
\beq
\label{3.1.36e}
S = (m_D^*)^{-1} K^*m_L^{diag}
\eeq

From inspection of the precedent equations, one sees that (\ref{3.1.35e},\ref{3.1.36e}) are relations between the mass basis quantities ($K, R, S, T, m_L^{diag}, M_R^{diag}$) and the current basis matrices $m_D, M_R^{diag}$, since $M_R$ is diagonalized and appears in both bases.
Eliminating $m_D$, one finds an exact relation between quantities in the mass basis :
\beq
\label{3.1.37e}
M_R^{diag}T^{-1}S = (R^*)^{-1}K^*m_L^{diag}
\eeq

\underline {The matrices $(K, R, S, T)$ for $m_D << M_R$}

\vskip 3truemm

If $m_D << M_R$, one has the order of magnitude
\beq
\label{3.1.38e}
R \sim S \sim O\left({m_D \over M_R}\right) 
\eeq

Neglecting in equations (\ref{3.1.21e}-\ref{3.1.26e}) of the Appendix the terms of $O\left({m_D^2 \over M_R^2}\right)$ one gets the {\it approximate unitarity conditions}
\beq
\label{3.1.39e}
K K^\dagger \simeq K^\dagger K \simeq 1 
\eeq
\beq
\label{3.1.40e}
T T^\dagger \simeq T^\dagger T \simeq 1 
\eeq

Moreover, from (\ref{3.1.39e},\ref{3.1.40e}), both equations (\ref{3.1.23e}) and (\ref{3.1.26e}) imply the same approximate relation between $R$ and $S$ 
\beq
\label{3.1.41e}
R \simeq - K S^\dagger T
\eeq

In conclusion, in the present approximation one gets two unitary matrices $K$ and $T$  (\ref{3.1.39e},\ref{3.1.40e}) and the matrix $R$ given in terms of $(K, T, S)$ by (\ref{3.1.41e}).\par

On the other hand, neglecting terms of $O\left({m_D^2 \over M_R^2}\right)$ in (\ref{3.1.30e}-\ref{3.1.32e}), one gets
\beq
\label{3.1.45e}
Km_L^{diag}K^t + RM_R^{diag}R^t = 0 \ \ \ \ \  
\eeq 
\beq
\label{3.1.46e}
TM_R^{diag}T^t \simeq M_R^{diag} \qquad \qquad \ \ \ \
\eeq
\beq
\label{3.1.47e}
RM_R^{diag}T^t \simeq m_D \qquad \qquad \qquad 
\eeq

\noindent Eqn. (\ref{3.1.46e}) implies 
\beq
\label{3.1.48e}
T \simeq 1
\eeq

\noindent Notice that (\ref{3.1.47e}) is identical to the relation (\ref{3.1.35e}) obtained above. On the other hand, combining (\ref{3.1.41e}) with the exact relation (\ref{3.1.37e}) one consistently obtains obtains (\ref{3.1.45e}).\par

One can see that (\ref{3.1.41e}) gives just the seesaw formula. From (\ref{3.1.35e},\ref{3.1.36e},\ref{3.1.48e}), eqn. (\ref{3.1.41e}) implies, after some algebra
\beq
\label{3.1.42e}
K m_L^{diag} K^t \simeq - m_D (M_R^{diag})^{-1} m_D^t 
\eeq

\noindent and from the general complex symmetric matrix $m_L$, 
\beq
\label{3.1.43e}
m_L = K m_L^{diag} K^t 
\eeq

\noindent one gets the seesaw formula in the ESM :
\beq
\label{3.1.44e}
\fbox{ $m_L \simeq -m_D(M_R^{diag})^{-1}m_D^t$ } 
\eeq

\noindent We see that $K$ is the mixing matrix for light neutrinos, that appears in (\ref{3.1.11e}) in the basis in which $m_e$ is diagonal.\par
On the other hand, relation (\ref{3.1.35e}) or (\ref{3.1.47e}), together with (\ref{3.1.48e}), implies 
\beq
\label{3.1.49e}
R \simeq m_D (M_R^{diag})^{-1}
\eeq

\noindent and using the seesaw formula (\ref{3.1.44e}), relation (\ref{3.1.36e}) becomes
\beq
\label{3.1.49bise}
S = - (M_R^{diag})^{-1} m_D^\dagger K
\eeq

\noindent in consistency with (\ref{3.1.41e}).

The whole set $K$, $R$, $S$, $T$ has 12(6) parameters, implying from (\ref{3.1.48e}) that $K$, $R$ and $S$ have 12(6) independent parameters. Since according to (\ref{3.1.49bise}) the matrix $S$ is not independent, the matrices $K, R$ that appear in the interaction Lagrangian (\ref{3.1.11e}), have together 12(6) parameters. From (\ref{3.1.49e}) and the 15(6) number of parameters of $m_D$, we see that $R$ will have 12(6) parameters. Since $K$ is unitary in the present approximation, we can choose 6(3) independent parameters {\it within} $R$ to provide the unitary matrix $K$ with 6(3) parameters, the physically relevant PMNS structure. Then $R$ will have other extra 6(3) parameters. However, other solutions are allowed, since $K$ is unitary, not necessarily of the PMNS type.\par

\subsubsection{Summary of the parameter counting in the mass basis}

In the mass basis, parameter counting in the physically relevant case is : 12(6) parameters from {\it both} the complex matrices $K, R$ (among these, 6(3) parameters from the PMNS-like matrix K) + 3(0) parameters from $M_R^{diag}$ + 3(0) parameters from $m_L^{diag}$ + 3(0) parameters from $m_e^{diag}$ = 21(6), the same counting as in the current basis. \par 
The more constrained condition $m_D << M_R$ provides a particular case : $R$ has 12(6) parameters, among which one has to choose the 6(3) parameters of the PMNS matrix $K$.

\subsection{\bf Left-Right Model}

\vskip 5.0 truemm

Let us start from the Lagrangian (\ref{2.17e}) of the LRM. At this stage $M_R$ is complex symmetric with 9(3) parameters. We rewrite (\ref{2.17e}) under the form
$${\cal{L}}_m = {1 \over 2} \left(\overline{\nu_L}, \ \overline{(N_R)^c} \right) {\cal{M}} \left(\begin{array}{c}
       (\nu_L)^c \\
       N_R\\
        \end{array}
        \right) + \overline{e}_L m_e^{diag} e_R + h.c. \qquad \ $$

$${\cal{L}}_w = \left(\overline{\nu_L}, \ \overline{(N_R)^c} \right) \gamma_\mu \left(\begin{array}{cc}
       1 & 0 \\
        0 & 0 \\
  	\end{array}
        \right) \left(\begin{array}{c}
       e_L \\
       e_L\\
        \end{array}\right)W_L^\mu \qquad \qquad \qquad \qquad \qquad $$
\beq
\label{3.2.1e}
+\ (\overline{N_R}, \ \overline{(\nu_L)^c}) \gamma_\mu \left(\begin{array}{cc}
       1 & 0 \\
        0 & 0 \\
  	\end{array}
        \right)\left(\begin{array}{c}
       e_R \\
       e_R\\
        \end{array}\right) W_R^\mu\ +\ h.c.  \qquad \qquad  
\eeq

\noindent where
\beq
\label{3.2.2e}
\cal{M} = \left(
        \begin{array}{ccc}
       0 & m_D \\
        m_D^t & M_R \\
  	\end{array}
        \right) 
\eeq

\noindent Unlike the case of the ESM, the complex symmetric block $M_R$ is not diagonalized, it has 9(3) parameters since three phases have been rotated away. 

Using the unitary matrix $V$ (\ref{3.1.3e}-\ref{3.1.6e}),
$$\left(\overline{\nu_L}, \ \overline{(N_R)^c} \right) \to \left(\overline{\nu_L}, \ \overline{(N_R)^c} \right)V^\dagger \qquad \qquad \qquad \qquad \ \ \ $$
\beq
\label{3.2.3e}
\left(\overline{N_R}, \ \overline{(\nu_L)^c}\right) \to \left(\overline{N_R}, \ \overline{(\nu_L)^c}\right) \left(\begin{array}{cc}
       0 & 1 \\
        1 & 0 \\
  	\end{array}
        \right) V^t \left(\begin{array}{cc}
       0 & 1 \\
        1 & 0 \\
  	\end{array}
        \right)
\eeq

\noindent we obtain the following Lagrangian in the mass basis
$${\cal{L}}_m = {1 \over 2}\ \overline{\nu_L}m_L^{diag} (\nu_L)^c + {1 \over 2}\ \overline{(N_R)^c}M_R^{diag} N_R  + \overline{e}_L m_e^{diag} e_R + h.c. \qquad \qquad  \qquad \qquad $$
\beq
\label{3.2.5e}
\fbox{ ${\cal{L}}_w = \left(\overline{\nu_L}K^\dagger + \overline{(N_R)^c}R^\dagger \right) \gamma_\mu e_L W_L^\mu +\left(\overline{N_R}T^t +\overline{(\nu_L)^c}S^t \right) \gamma_\mu e_R W_R^\mu + h.c.$ } \qquad    
\eeq

The $3 \times 3$ matrices $K$ and $R$ enter in the left sector, while $T$ and $S$ enter in the right sector, in a symmetric way. A formula of similar structure to eqn. (\ref{3.2.5e}) follow from the results of ref. \cite{LS-1}, that uses however a quite different notation.\par

It is important to point out that the terms dependent on $K$ and $T$ are lepton number conserving $\Delta L = 0$, while those that depend on $R$ and $S$ are lepton number violating $\Delta L = 2$.\par 

\vskip 3truemm

\subsubsection {The matrices $K$, $R$, $S$, $T$ in the Left-Right Model}

Particularizing (\ref{3.1.27e}-\ref{3.1.29e}) of the Appendix to the case (\ref{3.2.2e}), we have :
\beq
\label{3.2.16e}
Km_L^{diag}K^t + RM_R^{diag}R^t = 0 \ \ \ \  
\eeq 
\beq
\label{3.2.17e}
Sm_L^{diag}S^t + TM_R^{diag}T^t = M_R \ \ \
\eeq
\beq
\label{3.2.18e}
Km_L^{diag}S^t + RM_R^{diag}T^t = m_D \ \ 
\eeq

Considering for the moment only the unitarity of the full matrix $V$ (\ref{3.1.6e}), that has 36(21) parameters, the number of independent parameters in the l.h.s. of the precedent equations will be 36(21) from $(K, R, S, T)$ + 3(0) + from $m_L^{diag}$ + 3(0) from $M_R^{diag}$ = 42(21) parameters.\par

The complex symmetric matrix equation (\ref{3.2.16e}) gives 12(6) constraints. On the other hand, $M_R$ in the r.h.s. of (\ref{3.2.17e}) has 9(3) free parameters, and this equation implies 12(6) - 9(3) = 3(3) constraints. Finally, since $m_D$ is a general complex matrix, with 18(9) free parameters, eq. (\ref{3.2.18e}) does not give any constraint. This gives a total of 12(6) + 3(3) = 15(9) constraints. Therefore one has 42(21) - 15(9) = 27(12) independent parameters. Adding the 3(0) eigenvalues of $m_e^{diag}$, not counted up to now, one gets 27(12) + 3(0) = 30(12) parameters, the same result as in the current basis.\par 

Moreover, substracting from this total number of 30(12) parameters the 9(0) mass eigenvalues $m_e^{diag}$, $m_L^{diag}$ and $M_R^{diag}$, we see that the set of matrices $(K, R, S, T)$, that appear in the interaction term (\ref{3.2.5e}), have a total of 21(12) parameters.\par

\vskip 3truemm

In the $SU(2)_L \times SU(2)_R \times U(1)$ Model one obtains also the \underline {exact relations} between the matrices $K$, $R$, $S$, $T$ given above by eqns. (\ref{3.1.33e}-\ref{3.1.37e}).

\vskip 3.0 truemm

\underline{The matrices $(K, R, S, T)$ for $m_D << M_R$}

\vskip 3.0 truemm

The relations given above within the \underline{approximation $m_D << M_R$} (\ref{3.1.39e}-\ref{3.1.41e}) for the ESM also hold in the LR model.\par
Let us rewrite eqns. (\ref{3.2.16e}-\ref{3.2.18e}) neglecting terms of $O\left({m_D^2 \over M_R^2}\right)$ :
\beq
\label{3.2.31e}
Km_L^{diag}K^t + RM_R^{diag}R^t = 0 \ \ \ \  
\eeq 
\beq
\label{3.2.32e}
TM_R^{diag}T^t \simeq M_R \qquad \qquad \ \ \ \
\eeq
\beq
\label{3.2.33e}
RM_R^{diag}T^t \simeq m_D \qquad \qquad \ \ \ \ 
\eeq

Eqn. (\ref{3.2.33e}) is the above obtained exact relation (\ref{3.1.35e}) if one neglects in the latter higher order terms. This means that in (\ref{3.2.18e}), the first term of the l.h.s., that is of $O(m_D^3/M_R^2)$, is compensated by higher order terms in the second term $RM_R^{diag}T^t$. On the other hand, combining (\ref{3.1.41e}) with the exact relation (\ref{3.1.37e}), one consistently obtains the exact relation (\ref{3.2.31e}).\par

According to (\ref{3.1.40e}) and (\ref{3.2.32e}), the matrix $T$ is the unitary mixing matrix of right-handed neutrinos, for which we can take 6(3) parameters, i.e. a matrix of the PMNS type.
Eqn. (\ref{3.1.39e}) holds also in the LRM, and $K$ is the unitary mixing matrix of light left-handed neutrinos.\par

Since the whole set $K$, $R$, $S$ and $T$ has 21(12) parameters and the matrices $K$, $T$ have 6(3) parameters each, this implies that $R$ and $S$ can have together 9(6) extra independent parameters.\par

In the LR model, from relations (\ref{3.1.41e}) and (\ref{3.2.32e}) one obtains
\beq
\label{3.2.34e}
Km_L^{diag}K^t \simeq -m_DT^*(M_R^{diag})^{-1}T^\dagger m_D^t 
\eeq
\noindent i.e. the seesaw formula
\beq
\label{3.2.35e}
\fbox{ $m_L \simeq -m_DM_R^{-1}m_D^t$ } 
\eeq

\noindent where $M_R$ is not diagonalized, to be compared with the seesaw formula (\ref{3.1.44e}) in the case of the ESM.\par
Notice the important point that in Section 1 we have disregarded the possibility in the LRM of a Higgs triplet $\Delta_L$ that in principle could also contribute to the mass of the light neutrinos (see for example \cite{AF, BORAH}), so that formula (\ref{3.2.35e}) is only correct in the LRM if one neglects this type II seesaw contribution.

Equation (\ref{3.2.33e}) implies, using the approximate unitarity of $T$ : 
\beq
\label{3.2.36bise}
R \simeq m_D T^* (M_R^{diag})^{-1} 
\eeq

\noindent to be distinguished from (\ref{3.1.49e}), that holds in the ESM case. We see that in the LR case the PMNS matrix $T$ of the heavy neutrinos $T$ enters in the matrix $R$ and, on the other hand, the matrix $S$ satisfies relation (\ref{3.1.49bise}) that we found in the ESM.\par

\subsubsection{Summary of the parameter counting in the mass basis}

We have seen that the set of matrices $K$, $R$, $S$ and $T$ have together 21(12) parameters. Unlike the case of the ESM, in the LR model we have enough parameter space to accomodate two different PMNS matrices for $K$ and $T$, with 6(3) parameters each. Then, $R$ and $S$ can have together extra 9(6) parameters. However, this situation is not compulsory : there can be overlap between the parameters of all the four matrices $K$, $R$, $S$ and $T$.\par
In conclusion, the parameter counting in the physically interesting solution is as follows : 6(3) parameters from the PMNS-like unitary matrix $K$ + 6(3) parameters from the PMNS-like matrix $T$ + 9(6) extra parameters from the complex matrices $R$, $S$ + 3(0) from $M_R^{diag}$ + 3(0) from $m_L^{diag}$ + 3(0) from $m_e^{diag}$ = 30(12) parameters, the same number as in the current basis.\par 

\subsubsection{Possible observables in the Left-Right model}

The gauge bosons $W_L$ and $W_R$ are mixed in the Left-Right Model :
\beq
\label{3.2.6e}
W_L = \cos\zeta\ W_1 - \sin\zeta\ W_2, \qquad \qquad W_R = e^{i\omega}(\sin\zeta\ W_1 + \cos\zeta\ W_2)
\eeq

\noindent where $W_1$ and $W_2$ are mass eigenstates, and the mixing angle $\zeta$, in terms of the vacuum expectation values (\ref{4.3bise},\ref{4.6e}), is of the order \cite{LS-2}
\beq
\label{3.2.6bise} 
\zeta \simeq \pm {g_L \over g_R} {2 \mid k_1k_2 \mid \over \mid v_R \mid^2}
\eeq

From (\ref{3.2.5e}), it is interesting to write down the lightest mass vector boson $W_1$ couplings to leptons
\beq
\label{3.2.7e}
{\cal{L}}_w^{W_1} = \left[\cos\zeta\left(\overline{\nu_L}K^\dagger + \overline{(N_R)^c}R^\dagger \right) \gamma_\mu e_L + e^{i\omega} \sin\zeta \left(\overline{N_R}T^t +\overline{(\nu_L)^c}S^t \right) \gamma_\mu e_R\right] W_1^\mu + h.c.   
\eeq
\noindent Besides the $\sim \cos\zeta$ term that describes the processes $\Delta L = 0$ (\ref{3.1.11bise}) and  $\Delta L = 2$ (\ref{3.1.11tere}) as in the ESM case, the subleading term $\sim \sin\zeta$ describes the $\Delta L = 0$ process
\beq
\label{3.2.7bise}
N_R \to e_R W_1
\eeq

\noindent and another term describing the lepton-number violating decay $\Delta L = 2$ of the gauge boson
\beq
\label{3.2.8bise}
W_1 \to \overline{e}_R (\nu_L)^c  
\eeq

\noindent ($L(\overline{e}_R) = L((\nu_L)^c) = -L(e_R) = -L(\nu_L)$). However, the amplitude for this latter decay is very small, as we will see below.\par
On the other hand, the heavier vector boson $W_2$ couplings to leptons read :
\beq
\label{3.2.7tere}
{\cal{L}}_w^{W_2} = \left[-\sin\zeta\left(\overline{\nu_L}K^\dagger + \overline{(N_R)^c}R^\dagger \right) \gamma_\mu e_L + e^{i\omega} \cos\zeta \left(\overline{N_R}T^t +\overline{(\nu_L)^c}S^t \right) \gamma_\mu e_R\right] W_2^\mu + h.c.   
\eeq

\noindent Here, the subleading $\sim \sin\zeta$ term describes the $\Delta L = 0$ process
\beq
\label{3.2.9e}
W_2 \to \overline{e}_L \nu_L
\eeq

\noindent and the $\Delta L = 2$ transition, assuming the mass of $W_2$ heavier that the one of $N_R$ :
\beq
\label{3.2.10e}
W_2 \to \overline{e}_L (N_R)^c
\eeq

\noindent On the other hand, the leading  $\sim \cos\zeta$ term describes the process $\Delta L = 0$
\beq
\label{3.2.11e}
W_2 \to \overline{e}_R N_R
\eeq

\noindent and the $\Delta L = 2$ involving {\it light leptons} :
\beq
\label{3.2.12e}
W_2 \to \overline{e}_R (\nu_L)^c
\eeq

Of course, the phenomenological relevance of the $\Delta L = 2$ decay involving the $W_R$ gauge boson depends on its mass scale.\par

Concerning the possibility of physics of the LRM at relatively low energies, with observables at LHC scales, one should however take into account that there are severe constraints on such a low energy LRM. This point has been carefully studied in a detailed paper by Deshpande, Gunion, Kayser and Olness \cite{DGKO}, who have examined the relevant constraints : structure of the vacuum, limits on flavor-changing neutral currents, etc. The conclusion is that, although such a low energy LRM is not excluded, it is not natural in a straightforward way, and can only be formulated through some degree of fine-tuning.\par

If one assumes that the mass scale of the LRM is low, it makes sense to look at the LHC for lepton-number violation processes through the search of $pp \to \ell \ell jj$ topologies, where the two leptons are of the same charge (see for example the recent refs. \cite{HELO,ATLAS,CMS}).\par 
Indeed, using (\ref{3.2.7tere}) there is the possibility of the $\Delta L = 2$ process  
\beq
\label{3.2.12-1e}
W_2^+ \sim W_R^+ \to e_R^+ N_R \to e_R^+ e_L^+ W_L^- \to  e_R^+ e_L^+ jj  
\eeq

\noi where $W_L^-$ decays into two hadronic jets, the subscripts in $e_R$ and $e_L$ mean the couplings to $W_R$ and $W_L$, and we use the notation $(e_R)^c = e_R^+, (e_L)^c = e_L^+$. The decay chain (\ref{3.2.12-1e}) is the very interesting Keung-Senjanovi\' c process proposed long time ago \cite{KS} that tests, at the same time, the decay of the gauge boson $W_R$ and the Majorana character of the right handed neutrino $N_R$.\par

The PMNS mixing matrix $T$ of the heavy right-handed neutrinos $N_R$ controls the decay $W_R^+ \to e_R^+ N_R$. On the other hand, we see from formula (\ref{3.2.5e}) that the secondary decay $N_R \to e_L^+ W_L^-$ is controlled by the matrix $R\simeq m_D T^* (M_R^{diag})^{-1}$ (cf. (\ref{3.2.36bise})). Therefore, this latter decay is controlled by the Dirac mass \cite{KS} in the basis in which $M_R$ is diagonalized, $m'_D = m_D T^*$ (see below the leptogenesis part).\par

The decay chain (\ref{3.2.12-1e}) through $W_R^+ \to e_R^+ N_R \to e_R^+ e_L^+ W_L^-$ depends on both matrices $T$ and $R$.  
Let us suppose that, through the kinematics of the two jets in the decay $W_L^- \to jj$, one can reconstruct the $W_L^-$ boson. Then, the angular distribution of the three body decay $W_R^+ \to e_R^+ e_L^+ W_L^-$ will give information on the matrices $T$ and $R$.\par

The discussion of the observables in these decay chains depends on the assumed $N_R$ spectrum. One usually assumes that the gauge boson $W_R$ has a mass bigger or of the order of the heaviest $N_R$, that would correspond to a Yukawa coupling of $O(1)$, in analogy with the top quark. However, to simplify what follows, let us assume that all $N_{R_i}\ (i = 1, 2, 3)$ are lighter than the $W_R$. \par

From relation (\ref{3.2.36bise}) one can see that, in the limit of degenerate heavy neutrinos, summing over the three $N_{R_i}$, the amplitude for the process $W_R^+ \to e_R^+ e_L^+ W_L^-$ depends on the product $m_D^* T M_R^{-1}$, where $M_R$ is the {\it non-diagonalized} right-handed neutrino mass.\par

One relevant question is to ask whether one can measure the PMNS mixing matrix $T$. Then, although quite difficult, the $W_L$ could in principle be reconstructed through its decays into two jets $W_L \to jj$, and the different $N_{R_i}$ could be reconstructed as well through the decays $N_{R_i} \to e_L^+ W_L^-$.\par
Our starting point was the mass Lagrangian where the charged lepton part is diagonalized (\ref{3.2.1e}), and the final output was the interaction Lagrangian (\ref{3.2.5e}), where the decays $W_R^+ \to e_R^+ N_{R_i}\ (i = 1, 2, 3)$ depend on the PMNS matrix $T$. Considering the possibility of the three leptons $e_i\ (i = 1, 2, 3)$ of the Standard Model $e, \mu$, $\tau$, we see that through the rates of these decays, the {\it moduli} of all the matrix elements $T_{ij}$ are in principle accessible to experiment.   

\section{Representations of the Dirac mass matrix}

The Dirac mass matrix $m_D$ is a crucial input in neutrino physics, making the link between high and low energy. We review now some useful representations of  $m_D$. 

\subsection{Triangular parametrization}

An interesting representation of the Dirac mass matrix $m_D$ has been proposed by Branco et al. \cite{BMNR} :
\beq
\label{2.9bise}
m_D = Um_{\Delta}
\eeq 

\noindent where $U$ is a unitary matrix with 6(3) parameters of the PMNS form, although not identical to it, and $m_{\Delta}$ is a triangular matrix, with 3 vanishing off-diagonal elements, 3 real diagonal elements and 3 complex off-diagonal elements.\par 
The  factorization formula (\ref{2.9bise}) is usually called in Mathematics {\it QR Decomposition} of a complex square matrix $M$. In Mathematica notation \cite{WOLFRAM} \rm{QRDecomposition[M] gives the decomposition of a numerical complex matrix $M$ in terms of a unitary matrix $U$ and an upper triangular matrix $m_{\Delta}$, while \cite{HMP,BMNR} refers to a lower triangular matrix, although this is not an essential point. This decomposition can be numerically very useful for texture models of the matrix $m_D$, since it isolates $m_{\Delta}$, and hence the parameters that are relevant for leptogenesis.\par 
The counting of parameters for $m_D$ holds in (\ref{2.9bise}) : 15(6) parameters of $m_D$ = 6(3) parameters of $U$ + 9(3) parameters from the triangular matrix $m_{\Delta}$. Relation (\ref{2.9bise}) also holds if $m_D$ is general complex and $U$ a general unitary matrix : 18(9) parameters of $m_D$ = 9(6) parameters of $U$ + 9(3) parameters from the triangular matrix $m_{\Delta}$.
In the same way that 3 phases of $m_D$ can be rotated away by the transformation (\ref{2.7e}-\ref{2.9e}), and one can consistently rotate away 3 phases of the general unitary matrix $U$ \cite{BMNR}. \par
Relation (\ref{2.9bise}) is non-trivial. Indeed, because of the unitarity of $U$ we see that $m_D^\dagger m_D$ is given by
\beq
\label{3.9tere}
m_D^\dagger m_D = m_{\Delta}^\dagger m_{\Delta}
\eeq
and therefore the three CP phases of $m_{\Delta}$ control the amount of leptogenesis at high energies in the one-flavor approximation.\par

\subsubsection{Extended Standard Model}

With (\ref{2.9bise}), equation (\ref{3.1.49e}) obtained within the seesaw, writes
\beq
\label{3.1.52e}
R \simeq U m_{\Delta} (M_R^{diag})^{-1}
\eeq 

We have seen above that if we decide that $K$ is of the PMNS type with 6(3) parameters, then the parameters of $K$ have to be chosen among the ones of $R$. A solution satisfying this criterium is a Dirac mass matrix given by \cite{HMP} 
\beq
\label{3.1.53e}
m_D = K m_{\Delta}, \qquad \qquad \qquad R \simeq K m_{\Delta} (M_R^{diag})^{-1}
\eeq 

\noindent Besides its historical interest, this solution has the very nice feature of factorization of the Dirac mass matrix into two pieces, a low energy PMNS mixing matrix $K$ with 6(3) parameters, and a high energy mass matrix $m_{\Delta}$, that has 9(3) parameters and controls leptogenesis.\par

Another extreme case would be to assume that $U = 1$ \cite{BGJMRS,FT} that implies 
\beq
\label{3.1.53bise}
m_D = m_{\Delta}, \qquad \qquad \qquad R \simeq m_{\Delta} (M_R^{diag})^{-1}
\eeq 

\noindent This ansatz relates directly the CP-violating phase in leptogenesis and CP-violation at low energy in neutrino oscillations.\par 
However, there are many other solutions, since in all generality one can choose the parameters of $K$ among the ones of the product $m_D = U m_{\Delta}$.

\subsubsection{Left-Right Model}

\noindent Equation (\ref{3.2.36bise}), writes
\beq
\label{3.2.35bise}
R = U m_{\Delta} T^* (M_R^{diag})^{-1}
\eeq 

\noindent where we see that the matrix $T$, unlike the case of the ESM (\ref{3.1.52e}), enters in the definition of the matrix $R$, that controls leptogenesis.

\subsection{The orthogonal parametrization}

Another useful parametrization of $m_D$ has been proposed by Casas and Ibarra \cite{CI}.

\subsubsection{Extended Standard Model}

Starting from the seesaw formula (\ref{3.1.44e}) and diagonalizing $m_L$ by the PMNS matrix $K$  (\ref{3.1.43e}),
\beq
\label{3.2.36e}
m_L^{diag} = - K^\dagger m_D(M_R^{diag})^{-1}m_D^t K^* 
\eeq

\noindent As pointed out in \cite{CI}, this relation implies, 
\beq
\label{3.2.37e}
- (m_L^{diag})^{-1/2}K^\dagger m_D(M_R^{diag})^{-1/2}(M_R^{diag})^{-1/2}m_D^t K^*(m_L^{diag})^{-1/2} = 1 
\eeq

\noindent and therefore the matrix $i (m_L^{diag})^{-1/2}K^\dagger m_D(M_R^{diag})^{-1/2}$ is an orthogonal complex matrix $O$
\beq
\label{3.2.38e}
O = i(m_L^{diag})^{-1/2}K^\dagger m_D (M_R^{diag})^{-1/2}
\eeq

\noindent i.e. $O O^t = 1$. One finds the general expression for $m_D$ in terms of the matrix $O$
\beq
\label{3.2.39bise}
m_D = - i K (m_L^{diag})^{1/2} O (M_R^{diag})^{1/2}
\eeq

\noindent One can check from this expression that $m_D = K m_{\Delta}$ (\ref{3.1.53e}) is not the most general form for $m_D$ because $O$, being a general complex orthogonal matrix, the combination $-i (m_L^{diag})^{1/2} O (M_R^{diag})^{1/2}$ is not triangular in general.

The parametrization (\ref{3.2.39bise}) is very useful to analyze leptogenesis $CP$ asymmetries when taking flavor into account.

\subsubsection{Left-Right Model}

From eq. (\ref{3.2.34e}) one gets, instead of (\ref{3.2.37e})
\beq
\label{3.2.40e}
- (m_L^{diag})^{-1/2}K^\dagger m_D T^*(M_R^{diag})^{-1/2}(M_R^{diag})^{-1/2}T^\dagger m_D^t K^*(m_L^{diag})^{-1/2} = 1 
\eeq

\noindent that defines the orthogonal matrix
\beq
\label{3.2.41e}
O' = i(m_L^{diag})^{-1/2}K^\dagger m_D T^* (M_R^{diag})^{-1/2}
\eeq

\noindent and $m_D$ is now in the LRM
\beq
\label{3.2.39e}
m_D = - i K (m_L^{diag})^{1/2} O' (M_R^{diag})^{1/2} T^t
\eeq

\noindent that includes the PMNS mixing matrix $T$ of right-handed neutrinos.

\subsection{Relation between the triangular and orthogonal forms}

The orthogonal parametrization of the Dirac mass matrix $m_D$ appears to be powerful because it explicitly includes low energy quantities, the light neutrino eigenvalues $m_L^{diag}$ and the PMNS mixing matrix $K$ and, on the other hand, high energy quantities, the heavy right-handed neutrino eigenvalues $M_R^{diag}$ and an unknown orthogonal complex matrix $O$. One can write down the relation between both representations.\par
In the ESM, from relation (\ref{3.2.39bise}) one can write the QR decomposition of the matrix 
\beq
\label{3.2.41bise}
- i (m_L^{diag})^{1/2} O (M_R^{diag})^{1/2} =  Vm_\Delta
\eeq

\noindent where $V$ is another unitary matrix, and $m_\Delta$ a triangular matrix. We see therefore that the matrix $m_D$ has the form of the triangular parametrization (\ref{2.9bise}) $m_D = U m_\Delta$, with the PMNS matrix $K$ being a factorizable part of the unitary matrix $U$, namely $U = KV$. Therefore, although one can set $U = 1$, i.e. $V = K^{-1}$, and then the low energy phases are part of $m_\Delta$ and hence of leptogenesis, the natural solution seems to be that the PMNS matrix $K$ is a unitary factor of the matrix $U$, i.e. $U = KV$, $V$ being a unitary matrix. 

\section{Leptogenesis}

The gauge models that we consider conserve $B-L$. As nicely pointed out by Strumia \cite{STRUMIA}, the mere existence of sphalerons, that violate $B+L$ in the Standard Model at high temperature, suggests that baryogenesis can proceed via leptogenesis \cite{FY,DNN}. From (\ref{3.1.11e}) or (\ref{3.2.5e}), we see that lepton number is violated by the decays of heavy right-handed neutrinos, giving rise to a lepton asymmetry that is partially converted into a baryon asymmetry by the sphalerons. The out-of-equilibrium CP violating decays of heavy Majorana neutrinos, supplemented by sphaleron interactions, satisfy the three Sakharov criteria \cite{SAKHAROV} to obtain baryogenesis.\par

In this section we consider leptogenesis in the electroweak broken phase, coming from the $CP$ violating $\Delta L = 2$ decay $(N_R)^c \to e_LW_L$ in the Lagrangians (\ref{3.1.11e}) of the ESM and (\ref{3.2.5e}) of the LRM.\par 

The actual leptogenesis occurs at very high temperature, in the electroweak unbroken phase. The connection between cosmological $CP$ violation in the unbroken phase \cite{CRV} with a single massless Higgs doublet, and in the broken phase has been underlined by Branco et al. \cite{BMNR}. In the case of the Left-Right model, this connection is not clear a priori because the massless Higgs fields in the unbroken case belong to the bidoublet (\ref{4.1e}). As we emphasize below, this relation is worth to be investigated. For the moment, we are interested here in the possible differences between the ESM and the LRM in the broken phase, where the interaction Lagrangians (\ref{3.1.11e}) and (\ref{3.2.5e}) apply.

\subsection{One-flavor approximation}

\subsubsection{Extended Standard Model}

In this part on the ESM we reproduce the results of ref. \cite{BMNR}, with the aim of comparing below with the LRM.  The lepton number asymmetry from the decay of the $1$st, lightest heavy Majorana neutrino, in the broken electroweak phase and in the  {\it one-flavor approximation} is given by :
\beq
\label{4.1.1e}
\epsilon_1 = {g^2 \over M_W^2} {1 \over 16\pi}  {1 \over (R^\dagger R)_{11}} \sum_{k \not = 1} F(x_k)M_k^2Im[(R^\dagger R)_{1k}]^2
\eeq

\noindent since, from (\ref{3.1.11e}), the matrix $R$ is responsible for the transition $(N_R)^c \to e_L W_L$, or equivalently the decay $(N_R)^c \to e_L H$ above the phase transition. In eqn. (\ref{4.1.1e}) the function $F(x_k)$ reads 
\beq
\label{4.1.2e}
F(x_k) = \sqrt{x_k} \left[1 + (1+x_k) \ln \left({x_k \over 1+x_k} \right) + {1 \over 1-x_k} \right] \qquad \ \ \ \left( x_k = {M_k^2 \over M_1^2} \right)
\eeq\par

As poined out in ref. \cite{BMNR}, from (\ref{3.1.49e}) $R \simeq m_D (M_R^{diag})^{-1}$, that holds in the ESM for $m_D << M_R$, one gets the lepton number asymmetry in terms of the Dirac mass or, equivalently, in terms of the Yukawa couplings $m_D \over v$ in the unbroken phase :
\beq
\label{4.1.3e}
\epsilon_1 = {g^2 \over M_W^2} {1 \over 16\pi}  {1 \over (m_D^\dagger m_D)_{11}} \sum_{k \not = 1} F(x_k) \ Im[(m_D^\dagger m_D)_{1k}]^2
\eeq

\noindent While the expression of the lepton number asymmetry (\ref{4.1.1e}) depends only on quantities of the mass basis, namely on the matrices $R$, $M_R^{diag}$, expression (\ref{4.1.3e}) depends only on quantities of the current basis, since the matrix $M_R$ is diagonalized from the beginning in both bases.
Notice that, as exposed in \cite{BMNR}, expression (\ref{4.1.3e}) has a well-defined limit for the SM vacuum expectation value limit $v \to 0$, given in terms of Yukawa couplings corresponding to the decay  in the unbroken electroweak phase $(N_R)^c \to e_L H$ \cite{CRV}.\par 
In terms of the matrix $m_{\Delta}$ one gets
\beq
\label{4.1.4e}
\epsilon_1 = {g^2 \over M_W^2} {1 \over 16\pi}  {1 \over (m_{\Delta}^\dagger m_{\Delta})_{11}} \sum_{k \not = 1} F(x_k) \ Im[(m_{\Delta}^\dagger m_{\Delta})_{1k}]^2
\eeq  

\noindent that depends only on the three phases of $m_{\Delta}$.\par
On the other hand, in terms of the orthogonal matrix $O$ defined in (\ref{3.2.38e}) the CP asymmetry is given by
\beq
\label{4.1.5e}
\epsilon_1 = {g^2 \over M_W^2} {1 \over 16\pi} {1 \over M_1 \sum_i m_i\mid O_{i1} \mid^2} \sum_{k \not = 1} F(x_k)  \ M_1 M_k\ Im[\sum_{j} (m_j O_{j1})^2]
\eeq  

\subsubsection{\bf Left-Right Model}

In the LR model one has in principle two types of contributions to the light neutrino masses, through type I seesaw and type II seesaw, the latter arising from triplet Higgs exchange (see for example refs. \cite{AF,DGKO,BORAH}). As pointed out above, in this paper we consider only the contribution of type I seesaw.

In the LR case we have seen that the matrix responsible for the transitions $(N_R)^c \to e_L W_L$ is the matrix called also $R$ in the mass basis Lagrangian (\ref{3.2.5e}). Then, the lepton number asymmetry from the decay of the $1$st heavy Majorana neutrino, in the single flavor approximation, is given by the same formulas (\ref{4.1.1e},\ref{4.1.2e}).\par

In the LR model we have now $R$ given by (\ref{3.2.36bise}), that yields the lepton number asymmetry in terms of the Dirac mass and the mixing matrix $T$ of the heavy neutrinos :
\beq
\label{4.2.1e}
\epsilon_1 = {g^2 \over M_W^2} {1 \over 16\pi}  {1 \over (T^t m_D^\dagger m_D T^*)_{11}} \sum_{k \not = 1} F(x_k) \ Im[(T^t m_D^\dagger m_D T^*)_{1k}]^2
\eeq

\noindent In the LR model the lepton number asymmetry depends on the current basis matrix $m_D$ in (\ref{2.17e}) and also on the PMNS matrix $T$ of the heavy neutrinos. Consistently, the presence of the matrix $T$ appears in (\ref{4.2.1e}) because, to compute the decay rates $(N_1)^c \to e_L W_L$, one needs first to diagonalize the mass matrix $M_R = t v_R$ (\ref{4.6e}).\par
In other terms, the matrix $m_DT^* = m'_D$ is the Dirac mass matrix in the basis in which $M_R$ in (\ref{2.10e}) is diagonalized. In this basis the left-handed term of the interaction Lagrangian $\overline{\nu}_Le_L W_L$ remains diagonal, but the right-handed term $\overline{N}_Re_R W_R$ is not anymore.\par 

Expression (\ref{4.2.1e}) for the CP asymmetry in the electroweak broken phase follows from the $R$-term in the interaction Lagrangian (\ref{3.2.5e}), responsible for the decay $(N_R)^c \to e_LW_L$. This is the expression that has been used precisely to compute the leptogenesis CP asymmetry within LRM (see for example refs. \cite{BORAH,BBA}).\par

However, in the LRM the broken electroweak phase is more involved than in the ESM because there are two vacuum expectation values $k_1$ and $k_2$ (\ref{4.3bise}) that contribute to $m_D$ and to $M_W$, besides the possibility of a vacuum expectation value $v_L$ (not considered in subsection 1.1.2) that could also contribute to the $W_L$ mass.\par

In the unbroken electroweak phase, the Higgs bidoublet (\ref{4.1e}) would be massless, and one should consider both contributions $N_1 \to e\varphi_{1,2}$ to the leptogenesis asymmetry, with both Higgses $\varphi_{1,2}$ contributing to the loops needed to interfere with the tree diagram to obtain $CP$ violation. This situation reminds the one of the Standard Model with several Higgs doublets \cite{FO}. The relation between the $CP$ asymmetries in the broken and unbroken phases of the LRM deserves further investigation.\par

Since the matrix $m_D$ is general complex, so is $m_D T^*$ and we can write a decomposition in terms of another general unitary matrix $U'$ and another triangular matrix $m'_{\Delta}$ :
\beq
\label{4.2.2e}
m'_D = m_D T^* = U'm'_{\Delta}
\eeq

\noindent The lepton asymmetry writes
\beq
\label{4.2.3e}
\epsilon_1 = {g^2 \over M_W^2} {1 \over 16\pi}  {1 \over (m'^\dagger_\Delta m'_\Delta)_{11}} \sum_{k \not = 1} F(x_k) \ Im[(m'^\dagger_\Delta m'_\Delta)_{1k}]^2
\eeq

\noindent that now depends on the three CP phases of $m'_\Delta$.\par
On the other hand, notice that the interaction Lagrangian (\ref{3.2.5e}) contains also the $\Delta L = 2$ term $\overline{(\nu_L)^c}S^t e_R W_R$ that could give a contribution to the lepton asymmetry through the decay 
\beq
\label{4.2.3ebis}
W_R \to \overline{e}_R (\nu_L)^c 
\eeq
\noindent The masses $M_{W_R}$ and $M_i$ are both generated by the same Higgs triplet, and since one usually assumes that the Yukawa coupling of the heaviest neutrino $N_3$ is of $O(1)$, then $M_{W_R} >> M_1$ assuming a hierarchical spectrum for the heavy neutrinos. Hence, the lepton asymmetry generated by the decay of $W_R$ could be washed out and only the one due to the $N_1$ decays would survive. However, one should keep in mind in model building the possibility of leptogenesis through the decay (\ref{4.2.3ebis}).\par

\subsection{Flavored leptogenesis}

\subsubsection{Extended Standard Model}

A crucial progress in leptogenesis has been achieved by taking into account flavor \cite{BCST,ABADA-1,NNRR,ABADA-2}.
At high temperatures $T \geq 10^{12}$ GeV, all three $\tau, \mu$ and $e$ are out of equilibrium because their Yukawa couplings are weak compared to the temperature. In this regime, the one-flavor approximation can be applied since the different lepton flavors are undistinguishable.\par 
However, for "realistic" temperatures $T \simeq M_1$ such that $10^9 \leq T \leq 10^{12}$ GeV, the $\tau$ lepton doublet Yukawa coupling is large enough to be in termal equilibrium, while the $\mu$ and $e$ doublets are out of equilibrium. The net result is that the leptogenesis CP violation splits into two pieces, $\epsilon_\tau$ and $\epsilon_2 = \epsilon_\mu + \epsilon_e$, since the flavors $\mu$ and $e$ remain undistinguishable.
Then, in the range $10^9 \leq T \leq 10^{12}$ GeV, the final baryon asymmetry $Y_B$ is the sum of two contributions, given by the lepton CP asymmetries $\epsilon_\tau$ and $\epsilon_2$ affected by different wash-out factors $\eta_\tau$ and $\eta_2$ : $Y_B \propto \epsilon_\tau \eta_\tau + \epsilon_2 \eta_2$.
A recent updated flavor covariant description of flavor effects in leptogenesis can be found in ref. \cite{BDMPT}.\par   
The CP violating asymmetry for each flavour is given by the expression (see for example \cite{BFFNR}) :
$$\epsilon_{1\ell} = {g^2 \over M_W^2} {1 \over 16\pi}  {1 \over (m_D^\dagger m_D)_{11}} \sum_{k \not = 1} F(x_k) \ Im[(m_D^\dagger)_{1\ell}(m_D)_{\ell k} (m_D^\dagger m_D)_{1k}]^2$$
\beq
\label{4.2.4e}
\qquad \qquad +\ {g^2 \over M_W^2} {1 \over 16\pi}  {1 \over (m_D^\dagger m_D)_{11}} \sum_{k \not = 1} G(x_k) \ Im[(m_D^\dagger)_{1\ell}(m_D)_{\ell k} (m_D^\dagger m_D)_{k1}]^2
\eeq

\noindent where the second term corresponds to the lepton flavor violating but lepton number conserving self-energy diagram \cite{NNRR}. The function $F(x_k)$ is given by (\ref{4.1.2e}), and
\beq
\label{4.2.4bise}
G(x_k) = {1 \over 1-x_k}
\eeq

\noindent The second term in (\ref{4.2.4e}) vanishes when summing over $\ell$, while the first term gives the one-flavor approximation expression (\ref{4.1.3e}), because $\sum_\ell \epsilon_{1\ell} = \epsilon_1$. On the other hand, the second term in (\ref{4.2.4e}) is subleading if  one assumes $M_1 << M_2, M_3$.\par
The flavored wash-out factors read \cite{ABADA-2}
\beq
\label{4.2.4tere}
\eta_\ell = \eta\ {({m_D^\dagger})_{1\ell} {m_D}_{\ell 1} \over (m_D^\dagger m_D)_{11}}
\eeq

\noindent where $\eta$ is the wash-out factor in the single flavor approximation.\par 

Concerning the link between low energy CP violation in the PMNS mixing matrix and leptogenesis CP violation, the situation is quite different if flavor is taken into account \cite{ABADA-2}.
As an illustration, let us write the CP asymmetry $\epsilon_{1\ell}$, where the subindex 1 means decay of the lightest heavy Majorana neutrino $N_1$, by using the orthogonal parametrization (\ref{3.2.39bise}).
The flavor CP asymmetries $\epsilon_{1\ell}$ depend then on the low energy parameters, i.e. the light neutrino masses and the PMNS mixing matrix $K$. Assuming  $M_1 << M_2 < M_3$, one finds from (\ref{3.2.39bise}) and (\ref{4.2.4e}) the leptonic CP violation parameter $\epsilon_{1\ell}$ \cite{ABADA-2} :
\beq
\label{4.2.5e}
\epsilon_{1\ell} \simeq - {3 \over 32\pi}{g^2 \over M_W^2} {Im\left(\sum_{k,j}m_jm_k^{3/2}K^*_{\ell j}K_{\ell k} O^*_{j1} O^*_{k1}\right) \over {\sum_i m_i\mid O_{i1} \mid^2}}
\eeq

\subsubsection{Left-Right Model}

As we have seen in the LRM in the one-flavor approximation (formula (\ref{4.2.1e})), $m_D$ is replaced by $m_D T^*$, and the formula for the lepton asymmetry in this approximation is the same as in the Extended Standard model with the replacement $m_D \to m'_D = m_D T^*$ where $m'_D$ is the Dirac mass matrix in the basis in which the mass matrix $M_R$ is diagonalized.\par
Because of (\ref{3.2.41e}), formulas for the CP asymmetry (\ref{4.2.4e}) and the wash-out factor (\ref{4.2.4tere}) remain correct for the Left-Right model, with the replacement $m_D \to m'_D = m_D T^*$, where $m_D$ is given by (\ref{3.2.39e}), that has a complete left-right symmetry in the dependence on the mass eigenvalues $m_L^{diag}, M_R^{diag}$ as well as on the mixing matrices $K, T$. Then, the flavor asymmetry  has the same form (\ref{4.2.5e}), with the replacement $O \to O'$.

\section{Comparison between the Extended Standard Model and the Left-Right Model} 

We now summarize the comparison between the ESM and the LRM, as far as lepton mixing is concerned.\par

\vskip 0.2 truecm

(a) In the \underline{current basis} both models differ in the following way.\par 
In the ESM the Dirac matrix $m_D$ has 15(6) parameters because one can rotated away 3 phases and one can diagonalize the right-handed mass matrix $M_R$. One has finally a total of 21(6) parameters.\par
In the LRM one cannot diagonalize $M_R$ without changing the interaction Lagrangian. On the other hand, one cannot rotate away phases in both $m_D$ and in $M_R$, but only three phases in one of these matrices, that we have chosen to be $M_R$. Then, one is left with a general complex $m_D$ with 18(9) parameters and a complex symmetric $M_R$ with 9(3) parameters. With the $m_e$ mass eigenvalues, this gives a total of 30(12) parameters.\par
However, if in the LRM one diagonalizes $M_R$ from the start, the left-handed interaction term $\overline{\nu}_L \gamma_\mu e_L W_L^\mu$ remains diagonal, while the right-handed term $\overline{N}_R \gamma_\mu e_R W_R^\mu$ is modified. Also $m_D$ is modified to another Dirac mass term, that would eventually control leptogenesis. Therefore, as far as one considers the mass terms and the $W_L$ interation, one has the same number of parameters as in the ESM. For physics at low energy and also for leptogenesis, if the latter is attributed to the decays of the lightest right-handed heavy neutrino $N_1$, one can disregard the $W_R$ interaction term, that involves heavier degrees of freedom.

\vskip 0.2 truecm

(b) In the \underline{mass basis} in the ESM without approximations one has two mixing matrices $K$ and $R$ in the left sector, that have together 12(6) parameters. For $m_D << M_R$ one has a priori 12(6) parameters for the set of matrices $K, R$ (mixing in the left sector), and $S, T$ (mixing in the right sector). The mixing matrix of the left-handed neutrinos is approximately unitary and can be chosen to be of the PMNS type, with 6(3) parameters. The model constrains the mixing matrix of the right-handed neutrinos to be $T \simeq 1$, the matrix $R$ (\ref{3.1.49e}) has a total of to 12(6) parameters and $S$ is not independent because of relation\ (\ref{3.1.49bise}). The parameters of the PMNS mixing matrix for light neutrinos $K$ have to chosen among the ones of $R$. Adding the mass eigenvalues $m_L^{diag}, M_R^{diag}, m_e^{diag}$ one has a total of 21(6) parameters\par 

In the LRM in the mass basis one has more symmetry : two mixing matrices $K$, $R$ in the left sector and two $S$, $T$ in the right sector. These four matrices have {\it together} 21(12) parameters, that added to the mass eigenvalues $m_L^{diag}, M_R^{diag}, m_e^{diag}$ gives again a total of 30(12) parameters. In the approximation $m_D << M_R$ the mixing matrices $K$ (left sector) and $T$ (right sector) are unitary, and both can be chosen to be of the PMNS type, with 6(3) parameters each. This is different from the ESM for the right sector, where $T$ is trivial. This feature of the ESM seems unnatural, since physically one should expect a full PMNS matrix for the heavy right-handed neutrinos as well.\par

\vskip 0.2 truecm
 
(c) Adopting the \underline{decomposition $m_D = Um_{\Delta}$} ($U$ unitary and $m_{\Delta}$ triangular complex), in the ESM the matrix  $U$ has 6(3) parameters and $m_{\Delta}$ 9(3) parameters, corresponding to the 15(6) parameters of $m_D$. The natural solution is that the PMNS matrix $K$ is a unitary factor of the matrix $U$, namely $U = KV$, $V$ being also unitary. In the LRM the situation is somewhat different : $m_D$ is a general complex matrix with 18(9) parameters, $U$ is a {\it general} unitary matrix with 9(6) parameters and $m_{\Delta}$ has also 9(3) parameters. The Dirac mass matrix in the basis in which $M_R$ is diagonal (\ref{4.2.2e}) $m'_D = m_D T^*$ can be decomposed in the same way : $m'_D = U'm'_{\Delta}$.

\vskip 0.2 truecm

(d) Concerning the \underline{lepton asymmetry} relevant for leptogenesis, we find the following situation in both models.\par 
In the ESM, in the one-flavor approximation, the asymmetry is dependent on matrix elements of the matrices $R^\dagger R$ or $m_D^\dagger m_D$ or $m_\Delta^\dagger m_\Delta$, i.e. dependent on the 3 CP phases of $m_\Delta$. In the flavored case, the asymmetry (\ref{4.2.4e}) depends on the PMNS matrix $K$ and the three high energy phases of the orthogonal matrix $O$ (\ref{3.2.38e}).\par
In the LRM, in the one-flavor approximation, the lepton asymmetry is dependent on $R^\dagger R$ or $T^t m_D^\dagger m_D T^*$. Writing the product $m_D T^*$ as in (\ref{4.2.2e}), the asymmetry depends on the three CP phases of the triangular matrix $m'_\Delta$ through $m'^\dagger_\Delta m'_\Delta$. In the flavored case, the asymmetry depends on the three phases of the PMNS mixing matrix $K$ and on the three phases of $O'$ (\ref{3.2.41e}).\par
As far as model building is concerned, the situation is different in both schemes. As an example, imagine that one has a model for the Yukawas with some ansatz for $m_D$ and $M_R$. In the ESM, $M_R$ is diagonalized and $m_D$ is enough to compute the lepton asymmetry. In the LRM one needs to compute the matrix $T$ that diagonalizes $M_R$, in order to get $m'_D$.\par 

\vskip 0.2 truecm

(e) A possible identification between \underline{low energy phases and leptogenesis phases} is not possible in general. In the ESM one could imagine models in which the three CP phases of the light neutrinos mixing matrix $K$ are the same as the three phases of the triangular matrix $m_\Delta$, since one has to choose the parameters of $K$ among the ones of the matrix $R$ in the lepton asymmetry formula (\ref{4.1.1e}). In the LRM one could choose the three phases of $K$ to be the same as the ones of $m'_\Delta$ (\ref{4.2.2e}). 
\par
As to whether in general the leptogenesis CP asymmetry could depend on the low energy phases, in the flavored regime the usual argument that $\epsilon_{1\ell}$ in the ESM depend on the PMNS matrix $K$ and on the matrix $O$ (\ref{3.2.38e}) extends to the LRM with another orthogonal matrix $O'$ (\ref{3.2.41e}).

\vskip 0.2 truecm

(f) Relatively to the ESM, we have found that the LRM has some interesting new features :\par 
- The non-trivial PMNS mixing matrix $T$ of the heavy neutrinos enters in the quantitative estimation of decay branching ratios of heavy neutrinos $N_{R_i}$ to various final states.\par
- On the other hand, in the calculation of the leptogenesis CP asymmetries, the matrix $T$ is unobservable because the Dirac matrix that plays a role is now  (\ref{4.2.2e}) $m'_D = m_D T^*$, the Dirac matrix in the basis in which $M_R$ is diagonal.\par
- The term $\overline{(\nu_L)^c}S^t e_R W_R$ in (\ref{3.2.5e}) could give a contribution to the cosmological lepton asymmetry through the $\Delta L = 2$ lepton number violating decay to light leptons $W_R \to \overline{e}_R(\nu_L)^c$. 
As we have indicated above, this latter possibility seems unlikely in reasonable left-right models because $W_R$ is heavier than the lightest neutrino $N_1$. However, one should keep in mind this possibility in model building.\par
- Considering the $W_1, W_2$ basis, i.e. without neglecting $W_L-W_R$ mixing, we have seen in Section 3.2 that there is a term involving the lighter $W_1$ boson $\sim \sin\zeta\ \overline{(\nu_L)^c}S^t \gamma_\mu e_R\ W_1^\mu$, that allows for the subleading $\Delta L = 2$ lepton-number violating decay to light leptons $W_1 \to \overline{e}_R (\nu_L)^c$. 

\section{Extension to Pati-Salam and $SO(10)$} 

One can extend the precedent considerations to other left-right gauge models like the Pati-Salam gauge theory $SU(4)_C \times SU(2)_L \times SU(2)_R$ \cite{PS} or $SO(10)$ \cite{SO10}.\par
We can consider first each of these models in the current basis, with general mass terms determined only by the Dirac or Majorana character of the fermions, and perform the counting of the $CP$ conserving and $CP$ violating free parameters. In a second step, one can diagonalize the mass matrices and obtain mixing in the interaction terms and, in a third step, {\it switch on the Higgs sector} of each theory and see how, according to the different hypothesis on this sector, the predictive power of each scheme is improved. Of course, with the most general Higgs structure for each model, one populates the general parameter space of the mass terms obtained by imposing only Lorentz invariance.\par
Moreover, since in these theories leptons are related to quarks, lepton mixing in the Dirac mass term will be related to quark mixing, at least for some Higgs structures. This feature is interesting in view of increasing the predictive power of $SO(10)$ for leptogenesis, and has been used more or less quantitatively in the literature.\par

Let us give some details for the Pati-Salam model and for $SO(10)$. Consider first the general mass Lagrangian consistent with Lorentz invariance of Dirac and Majorana mass terms
\beq
\label{7.1e}
{\cal{L}}_m = \overline{\nu}_L m_D N_R + {1 \over 2}\ \overline{(N_R)^c}M_RN_R + \overline{e}_L m_e e_R + \overline{u}_L m_u u_R + \overline{d}_L m_d d_R + h.c.
\eeq

For the moment the matrices $m_D$, $m_e$, $m_u$ and $m_d$ are general complex with 18(9) parameters each and $M_R$ is a general complex symmetric matrix with 12(6) parameters. This gives a priori a total of 84(42) parameters, while in the lepton sector one has 18(9) (from $m_D$) + 18(9) (from $m_e$) + 12(6) (from $M_R$) = 48(24) parameters.\par 

In the Pati-Salam model and in $SO(10)$, the interaction Lagrangian has the general form 
\beq
\label{7.1-1e} 
 {\cal{L}}_{int} = {\cal{L}}_w  + {\cal{L}}_x
\eeq

\noindent where one has in both models, keeping only the interesting flavor-changing terms :
\beq
\label{7.1-1e} 
{\cal{L}}_w = \overline{e}_L \gamma_\mu \nu_L W_L^\mu +  \overline{e}_R \gamma_\mu N_R W_R^\mu + \overline{d}_L \gamma_\mu u_L W_L^\mu + \overline{d}_R \gamma_\mu u_R W_R^\mu + h.c. \ \  
\eeq

The extra interaction term writes in the Pati-Salam model :
\beq
\label{7.1-2e}
{\cal{L}}_x^{PS} = \overline{e}_L \gamma_\mu d_L X_L^\mu + \overline{e}_R \gamma_\mu d_R X_R^\mu + \overline{\nu}_L \gamma_\mu u_L X_L^\mu + \overline{N}_R \gamma_\mu u_R X_R^\mu + h.c. \ \ \  
\eeq

\noindent where the colored gauge bosons have charges $|Q(X_L)| = |Q(X_R)| = {2 \over 3}$.\par
In $SO(10)$ one has \cite{MACHACEK,GLYOPR} :
$${\cal{L}}_x^{SO(10)} = \left[\epsilon^{ijk} \overline{(u^i_R)^c} \gamma_\mu u^j_L + \overline{d^k_L} \gamma_\mu (e_R)^c - \overline{e_L} \gamma_\mu (d_R^k)^c\right]X^{k\mu} \qquad \qquad \qquad \qquad$$
$$\qquad \qquad \ \ + \left[\epsilon^{ijk} \overline{(u^i_R)^c} \gamma_\mu d^j_L + \overline{\nu_L} \gamma_\mu (d_R^k)^c - \overline{u_L^k}  \gamma_\mu (e_R)^c\right]Y^{k\mu} \qquad \qquad \qquad \qquad \qquad$$
$$\qquad \qquad \ \ \ + \left[\epsilon^{ijk} \overline{(d^i_R)^c} \gamma_\mu u^j_L + \overline{e_L} \gamma_\mu (u_R^k)^c - \overline{d_L^k}  \gamma_\mu (N_R)^c\right]Y'^{k\mu} \qquad \qquad \qquad \qquad \qquad$$ 
$$\qquad \qquad \ \ \ + \left[\epsilon^{ijk} \overline{(d^i_R)^c} \gamma_\mu d^j_L + \overline{\nu_L} \gamma_\mu (u_R^k)^c - \overline{u^k_L} \gamma_\mu (N_R)^c\right]X_{D}^{k\mu} \qquad \qquad \qquad \qquad \qquad$$
\beq
\label{7.1-3e}
\qquad \qquad \ \ + \left[\overline{\nu_L} \gamma_\mu u^k_L + \overline{e_L} \gamma_\mu d_L^k - \overline{(d_R^k)^c} \gamma_\mu (e_R)^c - \overline{(u_R^k)^c} \gamma_\mu (N_R)^c\right]S^{k\mu} + h.c. \ \ \  
\eeq

\noindent where $i, j, k$ are color indices and the colored gauge bosons $X, Y, Y', X_D, S$ have the charges : $|Q(X)| = {4 \over 3}, |Q(Y)| = |Q(Y')| = {1 \over 3}, |Q(X_D)| = |Q(S)| = {2 \over 3}$.\par
Let us see how many parameters can be rotated away in both models. 
Analogously to the LRM, one can diagonalize $m_e$ and absorb 3 phases in $M_R$ in (\ref{7.1e}) while keeping ${\cal{L}}_w$ (\ref{7.1-1e}) invariant. However, as it is obvious from (\ref{7.1-2e},\ref{7.1-3e}), ${\cal{L}}_x$ is changed under these transformations. In the pure lepton sector, leaving aside the quark-lepton terms in ${\cal{L}}_x$, the starting point for the diagonalization of the mass terms is the same as in the LRM (\ref{2.17e}), with 30(12) parameters in $m_D$, $M_R$ and $m_e^{diag}$. 
Diagonalizing (\ref{7.1e}) one gets the flavor-changing mixing in the interaction Lagrangian ${\cal{L}}_w  + {\cal{L}}_x$.\par  
In the pure lepton sector our conclusions are the following. The diagonalization has the same form for $SU(2)_L \times SU(2)_R \times U(1)$, Pati-Salam and $SO(10)$ models. Separately, the $3 \times 3$ matrices $K$ and $R$ enter in the left sector, while the $3 \times 3$ matrices $T$ and $S$ enter in the right sector, like in the LRM, eqn. (\ref{3.2.5e}). In $SU(2)_L \times SU(2)_R \times U(1)$, Pati-Salam and $SO(10)$ models we have {\it in the lepton sector} the same counting of free parameters, i.e. 30 real parameters, among them 12 CP-violating phases.\par
 
Let us now make some remarks on masses and mixing in some particular cases in the interesting $SO(10)$ case. Let us look at the product 
\beq
\label{6.1e} 
{\bf 16} \times {\bf 16} = {\bf 10}_S + {\bf \overline{126}} _S+ {\bf 120}_A
\eeq

\noindent where ${\bf 10}+{\bf \overline{126}}$ is the symmetric part and {\bf 120} the antisymmetric part. The representatios ${\bf 10}$ and ${\bf 120}$ are real, ${\bf 126}$ is complex, and the Yukawa terms that can give mass to the fermions are
\beq
\label{6.2e} 
{\bf 16}_f \times {\bf 16}_f \times {\bf 10}_H = {\bf 1} +\ ...
\eeq
\beq
\label{6.3e} 
{\bf 16}_f \times {\bf 16}_f \times {\bf \overline{126}}_H = {\bf 1} +\ ...
\eeq
\beq
\label{6.4e} 
{\bf 16}_f \times {\bf 16}_f \times {\bf 120}_H = {\bf 1} +\ ...
\eeq

The Yukawa part of the Lagrangian reads
\beq
\label{6.7e} 
{\cal{L}}_Y = {\bf 16}_f\left(Y_{10}{\bf 10}_H + Y_{126}{\bf \overline{126}}_H + Y_{120}{\bf 120}_H\right){\bf 16}_f
\eeq

\noindent where a possible sum over Higgs representations and Yukawa coupling matrices in family space is implicit. After spontaneous symmetry breaking one gets the mass Lagrangian (see for example \cite{LUZIO})
$$m_d = v^d_{10}Y_{10} + v^d_{126}Y_{126} + v^d_{120}Y_{120} \ \ $$
$$m_u = v^u_{10}Y_{10} + v^u_{126}Y_{126} + v^u_{120}Y_{120} \ \ $$
\beq
\label{6.9e} 
m_e = v^d_{10}Y_{10} - 3 v^d_{126}Y_{126} + v^e_{120}Y_{120}
\eeq
$$m_D = v^u_{10}Y_{10} - 3 v^u_{126}Y_{126} + v^D_{120}Y_{120}$$
$$M_R = v^R_{126}Y_{126} \qquad \qquad \qquad  \qquad \ \ $$

\noindent where the Yukawa matrices $Y_{10}$ and $Y_{126}$ are complex symmetric, $Y_{120}$ is complex antisymmetric, and the $v$'s are Higgs vacuum expectation values. From the term (\ref{6.2e}) alone we obtain the well-known relations $m_e = m_d$ and $m_D = m_u$, while the term (\ref{6.3e}) alone would give the relations $m_e = -3m_d$ and $m_D = -3m_u$, \noindent and {\it no relation} from the term (\ref{6.4e}).\par
The vev's in (\ref{6.9e}) are in all generality complex numbers if we assume that $CP$ can be spontaneously broken (soft $CP$ violation). If $CP$ is not spontaneously broken the vevs are real and all $CP$ violation comes from the Yukawa couplings (hard $CP$ violation).\par
One could wonder how within SO(10) one can get the most general counting of parameters done above, i.e. 84(42) parameters for the whole mass sector (\ref{7.1e}), with 48(24) parameters in the lepton sector. As said above, this is simply achieved if all the representations ${\bf 10}_H, {\bf \overline{126}}_H, {\bf 120}_H$ in (\ref{6.9e}) are present and are different for each mass matrix, that becomes then completely general.\par 

An interesting particular case is to consider only the ${\bf 10}$ and ${\bf \overline{126}}$ representations in (\ref{6.9e}), with ${\bf 120}$ absent :
$$m_d = m^d_{10} + m^d_{126} \ \ $$
$$m_u = m^u_{10} + m^u_{126} \ \ $$
\beq
\label{6.10e} 
m_e = m^d_{10} - 3 m^d_{126}
\eeq
$$m_D = m^u_{10} - 3 m^u_{126}$$
$$M_R = m^R_{126} \qquad \ \ \ \ $$

\noindent In this situation, all mass matrices $m_u, m_d, m_D, m_e$ and $M_R$ are complex symmetric.\par

Let us count again the number of parameters under this hypothesis. The complex symmetric matrices $m^d_{10}, m^d_{126}, m^u_{10}, m^u_{126}, m^R_{126}$, have 12(6) parameters each, that gives a total number of 60(30) parameters, a reduction relatively to the 84(42) total number of parameters of the general case. 
One can diagonalize the complex symmetric matrices $m_d,... M_R$ with unitary matrices $V_d,... V_R$. 
Because of relations (\ref{6.10e}), the unitary matrices $V_e$, $V_D$, $V_R$ are in principle given in terms of $V_u$ and $V_d$ and mass eigenvalues. Notice that, as discussed in the mass basis for the pure lepton sector, we can adopt without loss of generality the basis in which $m_e = m_e^{diag}$. However, these relations give complicated equations between the elements of mixing matrices. Within this case of considering both ${\bf 10}$ and  ${\bf \overline{126}}$, it seems hard to find relations between the mixing matrices in the quark and the lepton sector, at least in a model-independent way.\par 

Let us consider two limiting cases: while the ${\bf \overline{126}}$ contributes to $M_R$, only the ${\bf 10}$ or only the ${\bf \overline{126}}$ contribute to $m_d$, $m_u$, $m_e$ and $m_D$.\par

From (\ref{6.10e}) we see that in both cases one has quark-lepton symmetry in the mixing matrices, i.e. a relation between the left-handed neutrino Dirac mixing matrix $V_L$, where $m_D = V_L^\dagger m_D^{diag} V_R$, and the CKM quark matrix 
\beq
\label{6.13e} 
V_L = V_uV_d^\dagger = V_{CKM}
\eeq

\noindent This relation has been often used in a number of phenomenological schemes \cite{AFS,BFO,BFFNR}. However, as it is well known, one needs both representations ${\bf 10}$ and ${\bf \overline{126}}$ to describe fermion masses in $SO(10)$ \cite{GJ,HRR}, and therefore we must conclude that there is a clash between  a good description of fermion masses and the one of obtaining quark-lepton symmetry in mixing.\par

Although the point of view of obtaining useful theoretical hints from $SO(10)$ on the eigenvalues and mixing of the Dirac neutrino mass matrix has been advanced in a number of works \cite{F,AFS,AF,BFO,BFFNR}, it is worth to point out that there could be an alternative philosophy concerning the Dirac mass matrix. Within the Left-Right Model, if the $W_R$ gauge boson and the lightest heavy neutrino $N_R$ are light enough, there is the interesting possibility of a complete determination of the Dirac mass matrix from the experimental study of $W_R$ and $N_R$ decays \cite{NST}. 
 
\section{Conclusions} 

We have examined the parameter counting and structure of CP conserving and CP violating lepton mixing in two gauge models in the electroweak broken phase, the Extended Standard Model - i.e. the Standard Model plus one right-handed heavy neutrino per generation -, and the Left-Right Model $SU(2)_L \times SU(2)_R \times U_{B-L}(1)$. We have used both the "current basis", in which the gauge interactions are diagonal, and the "mass basis", where the mass matrices are diagonal and mixing appears in the charged current gauge-fermion part of the Lagrangian. On the other hand, we have distinguished between results that are exact and results that hold within the approximation of Dirac masses that are small relatively to right-handed neutrino masses, $m_D << M_R$.\par

We think that it is worth to compare these two models. One reason is that, for simplicity, in the literature people usually discuss lepton mixing within the simple ESM, while actually have in mind left-right Grand Unified Theories like $SO(10)$, that naturally include heavy right-handed neutrinos. The simplest LR model that we study in this paper is a kind of prototype for these more involved LR theories.\par

Although the outline of the parameter counting and structure of lepton mixing is rather close in both schemes, there are differences between the two models. In particular, the Extended Standard Model can accomodate a PMNS mixing matrix $K$ for light neutrinos, but there is no room in parameter space for a mixing matrix $T$ for the heavy neutrinos, the mixing matrix being close to the identity.  On the other hand, as one could naturally expect, the Left-Right Model is consistent with PMNS mixing matrices for both light and heavy neutrinos. The lepton asymmetry relevant for leptogenesis depends, not only on the Dirac mass $m_D$, but also on the matrix $T$, that is non-trivial. But the lepton asymmetry is given in terms of the Dirac mass in the basis in which the right-handed heavy neutrino mass matrix is diagonal, while the interaction term in the right-handed sector is not diagonal anymore.\par

In the case of the LR model, the connection between the lepton $CP$ asymmetry in the electroweak broken phase, coming from the decay $(N_R)^c \to W_L e_L$ and its $CP$ conjugate, and the one in the unbroken phase coming from the decay above the phase transition $N_R \to e \varphi$, where $\varphi$ is the Higgs bidoublet, is an open problem worth to be investigated.\par

Mixing in the LRM contains new terms that involve $\Delta L = 2$ CP violating interactions involving the $W_R$ gauge bosons. Considering the $W_L-W_R$ mixing, there are interesting new possible $\Delta L = 2$ processes with {\it light leptons in the final state} : the subleading decay $W_1 \to \overline{e}_R (\nu_L)^c$ and the leading one $W_2 \to \overline{e}_R (\nu_L)^c$. As emphasized above, it is worth to keep in mind, in model building, the possibility of the latter as a contribution to leptogenesis. \par
 
We have extended these results to other LR theories, namely the Pati-Salam model $SU(4)_C \times SU(2)_L \times SU(2)_R$ and the grand unified model SO(10), for which we find that the structure of mixing in the lepton sector is, in the most general case, the same as in the Left-Right Model $SU(2)_L \times SU(2)_R \times U_{B-L}(1)$. The specification of the Higgs sector provides schemes that have more predictive power.\par

If one assumes both symmetric ${\bf 10}$ and ${\bf 126}$ Higgs representations, necessary to describe the quark mass spectrum, we emphasize that there is a clash between the description of this spectrum and the assumption that the left-handed Dirac mixing matrix is approximately given by the quark CKM matrix, as sometimes it is assumed in phenomenological models arguing naive quark-lepton symmetry.\par

Phenomenological analyses are usually done within these gauge models as $SO(10)$ supplemented by simplifying hypotheses that give tractable schemes. But one should keep in mind that the general parameter space can yield other possibilities concerning the description of the interesting observables.\par 

Concerning low energy observables, there are no differences between the Extended Standard model and the minimal Left-Right model at leading order in $m_D/M_R$. The cosmological baryon asymmetry via leptogenesis above the electroweak phase transition deserves however further investigation within the Left-Right model.

\vskip 1.0 truecm

{\Large \bf Appendix}

\vskip 0.5 truecm

 {\Large A general digression on the matrices $K$, $R$, $S$, $T$}}

\vskip 0.5 truecm

To count the number of independent parameters in each scheme, it is useful to consider the general case of diagonalization of a $6 \times 6$ complex symmetric matrix,
\beq
\label{3.1.20e}
\cal{M} = \left(
        \begin{array}{ccc}
       m_L & m_D \\
        m_D^t & M_R \\
  	\end{array}
        \right) 
\eeq

\noindent where $m_L$ and $M_R$ are $3 \times 3$ complex symmetric. In general, a $6 \times 6$ complex symmetric matrix has 42(21) real parameters.

Let us now diagonalize $\cal{M}$ with the unitary matrix $V$ (\ref{3.1.4e}-\ref{3.1.6e}). The unitarity condition $VV^\dagger = 1$ is an hermitian relation that implies 36(15) constraints. A general complex $6 \times 6$ matrix has 72(36) parameters. Therefore, because of these constraints, $V$ must have $72(36) - 36(15) = 36(21)$ parameters, consistent with the number of ${n(n-1) \over 2}$ angles and ${n(n+1) \over 2}$ phases of a $n \times n$ unitary matrix.  Since ${\cal{M}}^{diag}$ has 6(0) parameters, the r.h.s. of (\ref{3.1.4e}) has 36(21) (from $V$) + 6(0) =  42(21), in consistency with the counting of parameters of the matrix $\cal{M}$ (\ref{3.1.20e}).\par

The unitarity of the matrix $V$ (\ref{3.1.6e}) implies \cite{HMP,BMNR}
\beq
\label{3.1.21e}
K K^\dagger + R R^\dagger = 1
\eeq
\beq
\label{3.1.22e}
S S^\dagger + T T^\dagger = 1
\eeq
\beq
\label{3.1.23e}
K S^\dagger + R T^\dagger = 0 
\eeq 
\beq
\label{3.1.24e}
K^\dagger K + S^\dagger S = 1
\eeq
\beq
\label{3.1.25e}
R^\dagger R + T^\dagger T = 1
\eeq
\beq
\label{3.1.26e}
K^\dagger R + S^\dagger T = 0
\eeq

Let us do the exercise of counting again the number of parameters of the matrices $(K, R, S, T)$. If each of them were general complex, we would have for each 18(9) parameters, that gives for $(K, R, S, T)$ a total of 72(36) parameters. Relations (\ref{3.1.21e}) and (\ref{3.1.22e}) are hermitian, giving each 9(3) constraints, while (\ref{3.1.23e}) is general complex, giving 18(9) constraints. In total, we have again 9(3) + 9(3) + 18(9) = 36(15) constraints, and therefore, the set $(K, R, S, T)$ has 72(36) - 36(15) = 36(21) independent parameters, in agreement with the counting of independent parameters of the unitary matrix $V$.\par

On the other hand, the diagonalization of (\ref{3.1.20e}) reads
\beq
\label{3.1.27e}
Km_L^{diag}K^t + RM_R^{diag}R^t = m_L  
\eeq 
\beq
\label{3.1.28e}
Sm_L^{diag}S^t + TM_R^{diag}T^t = M_R  
\eeq
\beq
\label{3.1.29e}
Km_L^{diag}S^t + RM_R^{diag}T^t = m_D  
\eeq

Verifying again the counting of parameters, we have for the r.h.s. of (\ref{3.1.27e}-\ref{3.1.29e}), 12(6) + 12(6) + 18(9)  parameters from respectively $m_L$, $M_R$ and $m_D$. This gives a total of 42(21) independent parameters for the r.h.s., that is equal to the number of parameters of the l.h.s., 36(21) + 3(0) + 3(0)  from, respectively $(K, R, S, T)$, $m_L^{diag}$ and $M_R^{diag}$.

\vskip 0.5 truecm

\section*{Acknowledgements} \hspace*{\parindent} 
We are grateful to Dr. V. Tello for reminding us the Keung-Senjanovi\' c effect \cite{KS} and for pointing out a theoretical study about the interesting possibility of measuring the neutrino Dirac mass matrix within the Left-Right model \cite{NST}. We are also indebted to Dr. P. Bhupal Dev for calling our attention to a recent updated formulation of flavor effects in leptogenesis \cite{BDMPT}.

\end{document}